\documentclass[longauth]{aa}  

\usepackage{CJK}
\newcommand{\gkai}[1]{\begin{CJK*}{UTF8}{gkai}\raisebox{.1em}{(}#1\raisebox{.1em}{)}\end{CJK*}}

\usepackage[colorlinks=true,linkcolor=blue,citecolor=blue, urlcolor=blue]{hyperref}
\usepackage{natbib,hyperref}

\usepackage{graphicx}
\usepackage{txfonts}
\usepackage{placeins}
    \makeatletter
\def\@fnsymbol#1{\ensuremath{\ifcase#1\or \dagger\or \ddagger\or\mathsection\or \mathparagraph\or \|\or **\or \dagger\dagger\or \ddagger\ddagger \else\@ctrerr\fi}}
    \makeatother

\defcitealias{2023martin}{MS23}

\begin{document} 

   \title{The Pristine survey\thanks{Based on service-mode and visitor-mode observations made with the INT/IDS operated on the island of La Palma by the \textit{Isaac Newton} Group of Telescopes in the Spanish Observatorio del Roque de los Muchachos of the Instituto de Astrof\'isica de Canarias.}}
   \subtitle{XXV. The very metal-poor Galaxy: Chemodynamics through the follow-up of the Pristine-\textit{Gaia} synthetic catalogue\thanks{Table \ref{catalogue} released with this paper is only available in electronic form at the CDS via anonymous ftp to \href{https://cdsarc.cds.unistra.fr/}{cdsarc.u-strasbg.fr} (\href{ftp://130.79.128.5/}{130.79.128.5}) or via \href{http://cdsweb.u-strasbg.fr/cgi-bin/qcat?J/A+A/}{http://cdsweb.u-strasbg.fr/cgi-bin/qcat?J/A+A/}}}
   \titlerunning{The very metal-poor Galaxy}


   \author{Akshara Viswanathan\inst{\ref{kapteyn}}\thanks{Corresponding author \email{astroakshara97@gmail.com}}
          \and
          Zhen Yuan \gkai{袁珍}\inst{\ref{stra}}
          \and
          Anke Ardern-Arentsen\inst{\ref{cam}}
          \and
          Else Starkenburg\inst{\ref{kapteyn}}
          \and
          Nicolas F. Martin\inst{\ref{stra},\ref{mpia}}
          \and
          Kris~Youakim\inst{\ref{stock}}
          \and
          Rodrigo A. Ibata\inst{\ref{stra}}
          \and
          Federico Sestito\inst{\ref{uvic}}
          \and
          Tadafumi Matsuno\inst{\ref{heid}}
          \and
          Carlos Allende Prieto\inst{\ref{iac},\ref{laguna}}
          \and
          Freya~Barwell\inst{\ref{int},\ref{sheff}}
          \and
          Manuel Bayer\inst{\ref{kapteyn}}
          \and
          Amandine Doliva-Dolinsky\inst{\ref{tampa},\ref{dart},\ref{stra}}
          \and
          Emma Fern\'andez-Alvar\inst{\ref{iac},\ref{laguna}}
          \and
          Pablo~M.~Gal\'an-de~Anta\inst{\ref{int}}
          \and
          Kiran~Jhass\inst{\ref{int},\ref{sheff}}
          \and
          Nicolas Longeard\inst{\ref{epfl}}
          \and
          Jos\'e Mar\'ia Arroyo-Polonio\inst{\ref{iac},\ref{laguna}}
          \and
          Pol Massana\inst{\ref{int}}
          \and
          Martin~Montelius\inst{\ref{kapteyn}}
          \and
          Samuel~Rusterucci\inst{\ref{stra}}
          \and
          Judith Santos-Torres\inst{\ref{int},\ref{iac},\ref{laguna}}
          \and
          Guillaume F. Thomas\inst{\ref{iac},\ref{laguna}}
          \and
          Sara Vitali\inst{\ref{int},\ref{chile}}
          \and
          Wenbo~Wu\inst{\ref{china},\ref{iac},\ref{laguna}}
          \and
          Paige Yarker\inst{\ref{int}}
          \and
          Xianhao~Ye\inst{\ref{china},\ref{iac},\ref{laguna}}
          \and
          David S. Aguado\inst{\ref{iac},\ref{laguna}}
          \and
          Felipe Gran\inst{\ref{nice}}
          \and
          Julio Navarro\inst{\ref{uvic}}
          }

   \institute{Kapteyn Astronomical Institute, University of Groningen, Landleven 12, 9747 AD Groningen, The Netherlands\label{kapteyn}\and
   Universit\'e de Strasbourg, CNRS, Observatoire astronomique de Strasbourg, UMR 7550, F-67000 Strasbourg, France\label{stra}\and
   Institute of Astronomy, University of Cambridge, Madingley Road, Cambridge CB3 0HA, UK\label{cam}\and
   Max-Planck-Institut f\"ur Astronomie, K\"onigstuhl 17, D-69117 Heidelberg, Germany\label{mpia}\and
   Department of Astronomy, Stockholm University, AlbaNova University Centre, SE-106 91 Stockholm, Sweden\label{stock}\and
   Dept. of Physics and Astronomy, University of Victoria, P.O. Box 3055, STN CSC, Victoria BC V8W 3P6, Canada\label{uvic}\and
   Astronomisches Rechen-Institut, Zentrum f\"ur Astronomie der Universit\"at Heidelberg, M\"onchhofstra{\ss}e 12-14, 69120 Heidelberg, Germany\label{heid}\and
   Instituto de Astrof\'isica de Canarias, Calle V\'ia L\'actea s/n, 38206 La Laguna, Santa Cruz de Tenerife, Spain\label{iac}\and
   Universidad de La Laguna, Avda. Astrof\'isico Francisco S\'anchez, 38205 La Laguna, Santa Cruz de Tenerife, Spain\label{laguna}\and
   Isaac Newton Group of Telescopes, Apartado 321, 38700 Santa Cruz de la Palma, Tenerife, Spain\label{int}\and
   Department of Physics and Astronomy, University of Sheffield, Sheffield S3 7RH, UK\label{sheff}\and
   Department of Physics \& Astronomy, University of Tampa, 401 West Kennedy Boulevard, Tampa, FL 33606, USA\label{tampa}\and
   Department of Physics and Astronomy, Dartmouth College, Hanover, NH 03755, USA\label{dart}\and
   Institute of Physics, Laboratory of Astrophysics, Ecole Polytechnique F\'ed\'erale de Lausanne (EPFL), Observatoire de Sauverny, 1290 Versoix, Switzerland\label{epfl}\and
   N\'ucleo de Astronom\'ia, Facultad de Ingenier\'ia y Ciencias Universidad Diego Portales, Ej\'ercito 441, Santiago, Chile\label{chile}\and
   CAS Key Laboratory of Optical Astronomy, National Astronomical Observatories, Chinese Academy of Sciences, Beijing 100101, People's Republic of China\label{china}\and
   Universit\'e C\^ote d'Azur, Observatoire de la C\^ote d'Azur, CNRS, Laboratoire Lagrange, Nice, France\label{nice}}
   \date{Received 21 May 2024; accepted 23 July 2024}
    
    \abstract
   {The Pristine-\textit{Gaia} synthetic catalogue of reliable photometric metallicities makes use of spectrophotometric information from \textit{Gaia} DR3 XP spectra to calculate metallicity-sensitive \textit{CaHK} magnitudes, which in turn provides photometric metallicities for $\sim$30 million FGK stars using the Pristine survey model and the survey's training sample.}
   {We performed the first low- to medium-resolution spectroscopic follow-up of bright (G<15) and distant (upto 35 kpc) very and extremely metal-poor (V/EMP, [Fe/H]<-2.5) red giant branch stars from this catalogue--to evaluate the quality of the photometric metallicities and study the chemodynamics of these V/EMP stars.}
   {We used \textit{Isaac Newton} Telescope/Intermediate Dispersion Spectrograph (INT/IDS) observations centred around the calcium triplet region ideal for V/EMP stars for this spectroscopic follow-up.}
   {We find that 76\% of our stars indeed have [Fe/H] < -2.5 with these inferred spectroscopic metallicities, and only 3\% are outliers with [Fe/H] > -2.0. We report a success rate of 77\% and 38\% in finding stars with [Fe/H] < -2.5 and -3.0, respectively. This is a huge improvement compared to the literature in the selection of V/EMP stars based on photometric metallicities and will allow for 10,000--20,000 homogeneously analysed EMP stars using the WEAVE survey follow-up of Pristine EMP candidates. Using kinematics, we categorised 20\%, 46\%, and 34\% of the stars as being confined to the disc plane, or having inner and outer halo orbits, respectively. Based on their integrals-of-motion, we are able to associate these V/EMP stars with the metal-poor tail of the metallicity distribution functions of known accretion events such as the Gaia-Enceladus-Sausage, LMS-1/Wukong, Thamnos, Helmi streams, Sagittarius, Sequoia, and other retrograde mergers. For the stars that orbit close to the disc plane, we find that the prograde region with low vertical action is overdense with a significance of 4$\sigma$ compared to its retrograde counterpart.
   We also find three new (brightest) members of the most metal-poor stellar stream, C-19, one of which is 50$^\circ$ from the main body of the stream. This is the first member of C-19 found at positive height above the disc plane. Our measured mean metallicity, velocity dispersion, and stream width are consistent with the literature, but our results favour a slightly farther distance ($\sim$21.5 kpc) for the stream.}
   {With this work, we publish a catalogue (and 1D spectra) of 215 V/EMP stars from this first spectroscopic follow-up of the Pristine-\textit{Gaia} synthetic catalogue of photometric metallicities and showcase the power of chemokinematic analysis of bright and distant red giant stars in the V/EMP end.}

\maketitle

\section{Introduction}

The enduring presence of the most metal-poor stars within the Milky Way serves as an invaluable window to the early Universe and the pristine conditions in which these stars originated. These stellar relics are believed to have coalesced from material enriched by the initial generations of stars, offering crucial insights into the characteristics of their progenitors through their chemical compositions \citep{2005beers,2015frebel}. Furthermore, by studying the dynamical properties of these ancient stars, we gain valuable insights into the early formation processes of our Galaxy.

The metal-poor halo of the Milky Way has become the heart of several accreted structures, remnants of the Galaxy's tumultuous merger history. These structures include the Gaia-Enceladus-Sausage (GES), Sequoia, Thamnos, Helmi, LMS-1/Wukong, Cetus-Palca, and Sagittarius streams \citep{2018belokurov,2018helmi,2019barba,2019myeong,2019koppelman,2020yuan,2020naidu,2022thomas,1994ibata}, among others, each offering unique insights into past Galactic interactions. The recent discovery of numerous stellar streams further underscores the ongoing accretion events from dwarf and ultra-faint galaxies, as well as globular clusters, enriching the Galactic halo \citep{2022listreams,2021ibata,2023ibata,2022martin,2022martinb,2023viswanathan}. Additionally, recent studies suggest that a significant fraction of the halo stars may have formed \textit{in situ}, representing a blend of components including an $\alpha$-rich splashed hot thick disc and stars born in a primordial, hot, and disordered state \citep[e.g., ][]{2017bonaca,2018haywood,2019dimatteo,2020belokurov,2022belokurov}. 
This comprehensive understanding of the Milky Way's metal-poor population provides a foundation for unraveling the complex tapestry of Galactic formation and evolution.

The prevailing narrative across various cosmological simulations suggests that very metal-poor (VMP, [Fe/H] < -2.0) stars residing within the central regions of the Milky Way, including the bulge, represent some of the oldest stars in our Galaxy. These stars serve as invaluable tracers of the early Galactic assembly, offering critical insights into its formation history \citep{2005diemand,2010tumlinson,2017astarkenburg,2018elbadry}. On the observational front, numerous VMP stars have been scrutinised for their chemistry and kinematics, with particular emphasis on regions such as the bulge and the disc \citep{2014howes,2019lucey,2020arentsen}, providing essential data for understanding their origins.
The chemical characteristics of these stellar populations reveal a diverse assortment of objects that contributed to the formation of the inner Galaxy. Some stars exhibit chemical signatures indicative of formation in systems resembling ultra-faint dwarf galaxies, while others display traits consistent with birth within globular cluster-like environments \citep{2017schiavon}. Additionally, as pointed out earlier, there is evidence pointing towards a significant presence of \textit{in situ} metal-poor stars within the inner Galaxy \citep{2019conroy,2022belokurov,2023belokurov,2022rix,2024ardernarentsen}. This amalgamation of observational and theoretical findings paints a picture of the early stages of Galactic evolution, underscoring the complex interplay between various stellar populations and their environments.

Finding the most metal-deficient ([Fe/H] < -3.0) stars has long been recognised as a formidable challenge due to their exceptionally rare occurrence rate \citep{1981bond}. These stars are predominantly located in the Milky Way's halo component, and the ratio of halo to disc stars in the solar neighbourhood is approximately 10$^{-3}$ \citep{1980bahcall}. Moreover, the number of stars diminishes exponentially by a factor of roughly $\sim$10 or more for each dex decrease in metallicity \citep{1976hartwick}. In the solar neighbourhood, this translates to expectations of encountering one star with [Fe/H] = -3.0 among every 65,000 stars, and one star with [Fe/H] = -3.5 among 200,000 stars \citep{2020youakim}. To effectively address these challenges and advance our understanding of the old halo component with sufficient statistical power, efficient selection techniques are imperative.

Over the past four decades, numerous dedicated searches have been conducted to assemble large samples of the most metal-poor stars. Various techniques have been employed to identify these stars, including the pursuit of high proper motion stars exhibiting ultraviolet excesses \citep{1991ryan}, the detection of objects with diminished Ca II H \& K lines in extensive objective-prism surveys \citep{1985beers,2008christlieb}, and the utilisation of metallicity-sensitive (narrowband) photometry \citep{2014schlaufman,2017starkenburg,2018wolf,2019cenarro,2022almeida}.
Moreover, very and extremely metal-poor (V/EMP) stars have been identified in greater abundance through large-scale spectroscopic surveys such as the Sloan Digital Sky Survey \citep[SDSS,][]{2000york}, the Large sky Area Multi-Object fiber Spectroscopic Telescope \citep[LAMOST,][]{2006zhao,2018li}, the RAdial Velocity Experiment \citep[RAVE,][]{2006steinmetz}, the GALactic Archaeology with HERMES spectroscopic survey \citep[GALAH,][]{2021buder}, and the \textit{Gaia} Radial Velocity Spectrometer \citep[RVS,][]{2023recio,2022matsuno,2024viswanathan}. These photometric and spectroscopic surveys are often complemented by dedicated follow-up efforts at the VMP end \citep[e.g.][]{2019aguado, 2021yong, 2022li,2022witten}.

While astrometric data from the \textit{Gaia} mission \citep{2022gaia,2023gaia} has revealed significant stellar kinematic signatures attributed to earlier accretion events, there persists a crucial requirement for precise metallicities for many stars, particularly in the extremely metal-poor regime (EMP, [Fe/H] < -3.0). Such detailed metallicity measurements are essential for constructing a comprehensive chemodynamic framework, enabling a thorough understanding of the early structures that persist within the present Milky Way halo. 

In this work, our aim is to increase the amount of chemodynamically analysed bright (and distant) metal-poor red giant branch (RGB) stars. We used the Pristine-\textit{Gaia} synthetic catalogue of photometric metallicities to select V/EMP stars. These metallicities were estimated using the Gaia XP low-resolution spectrophotometry by determining the flux around the metallicity-sensitive narrowband Ca II H$\&$K lines (hereafter \textit{CaHK}), and running them with the Pristine survey model. We present the first results of our low- to medium-resolution follow-up spectroscopy of 215 stars in the Pristine-\textit{Gaia} synthetic catalogue to assess the performance of the survey's photometric pre-selection. We used Gaia's astrometry, distances, radial velocities (where available) and the follow-up spectra's radial velocities to calculate the dynamics, and we discuss their chemokinematic properties. Section \ref{2} summarises the target selection strategy for this spectroscopic follow-up program. Section \ref{3} presents the spectral analysis methods. Section \ref{4} presents the success rates of the photometric metallicities and discuss the chemokinematics of the V/EMP stars. Section \ref{5} discusses in detail the implications of disc-like V/EMP stars in the sample and new tentative members of the most metal-poor C-19 stellar stream. In section \ref{6}--we summarise the conlusions and outlook. 

\section{Target selection}\label{2}
In this section, we describe how the V/EMP stars were selected using the Pristine-\textit{Gaia} synthetic photometric metallicity catalogue. 
For a comprehensive overview of the Pristine survey, we refer to \cite{2017starkenburg}. 
Additionally, for detailed information on the Pristine-\textit{Gaia} synthetic catalogue and the Pristine data release 1, we refer to the work by \citet[][hereafter \citetalias{2023martin}]{2023martin}. 

\subsection{The Pristine survey and success rates}

The Pristine survey focuses on observing the sky of the northern hemisphere using the MegaCam wide-field imager installed on the Canada France Hawaii Telescope situated on Mauna Kea. 
This survey employs a narrowband filter centred on the metallicity-sensitive \textit{CaHK} in the near UV $(3968.5$ and $3933.7 \AA)$. 
When combined with Sloan Digital Sky Survey (SDSS) \textit{g, r, i} filters, or the \textit{Gaia} broadband BP-RP (blue and red photometric) and \textit{Gaia} G bands, this narrowband filter has been demonstrated to provide very reliable estimates of stellar metallicity and an exceptional tool for identifying V/EMP stars.

The Pristine photometry, in conjunction with SDSS broadband photometry, has paved the way for medium- and high-resolution spectroscopic follow-up studies. 
These investigations specifically target stars with the lowest metallicity estimates derived from Pristine \textit{CaHK} observations. 
Through dedicated medium- and high-resolution spectroscopic follow-up studies of Pristine-selected candidates for EMP stars, we have found strong evidence that the majority of these stars are indeed (very) metal-poor. 
Spectroscopic analyses of these candidates have confirmed that approximately 20\% of the EMP star candidates possess [Fe/H] < -3.0 and 70\% of the VMP star candidates are below [Fe/H] < -2.0 when carefully considering quality flags during target selection.
This conclusion is supported by the work of \citet{2017youakim}, and \citet{2019aguado} through medium-resolution follow-up and \citet{2017caffau}, \citet{2018starkenburg}, \citet{2019bonifacio}, \citet{2020caffau}, \citet{2020venn}, \citet{2021kielty}, \citet{2021lardo}, \citet{2022lucchesi}, and \citet{2023caffau} through high-resolution follow-up studies.
However, it is essential to account for potential variability in these stars and undertake thorough photometric quality cuts to ensure accurate characterisation of their metallicities \citep{2023lombardo}.

\subsection{Pristine-\textit{Gaia} synthetic catalogue of photometric metallicities}\label{2.2}

\begin{figure}
    \includegraphics[width=\columnwidth]{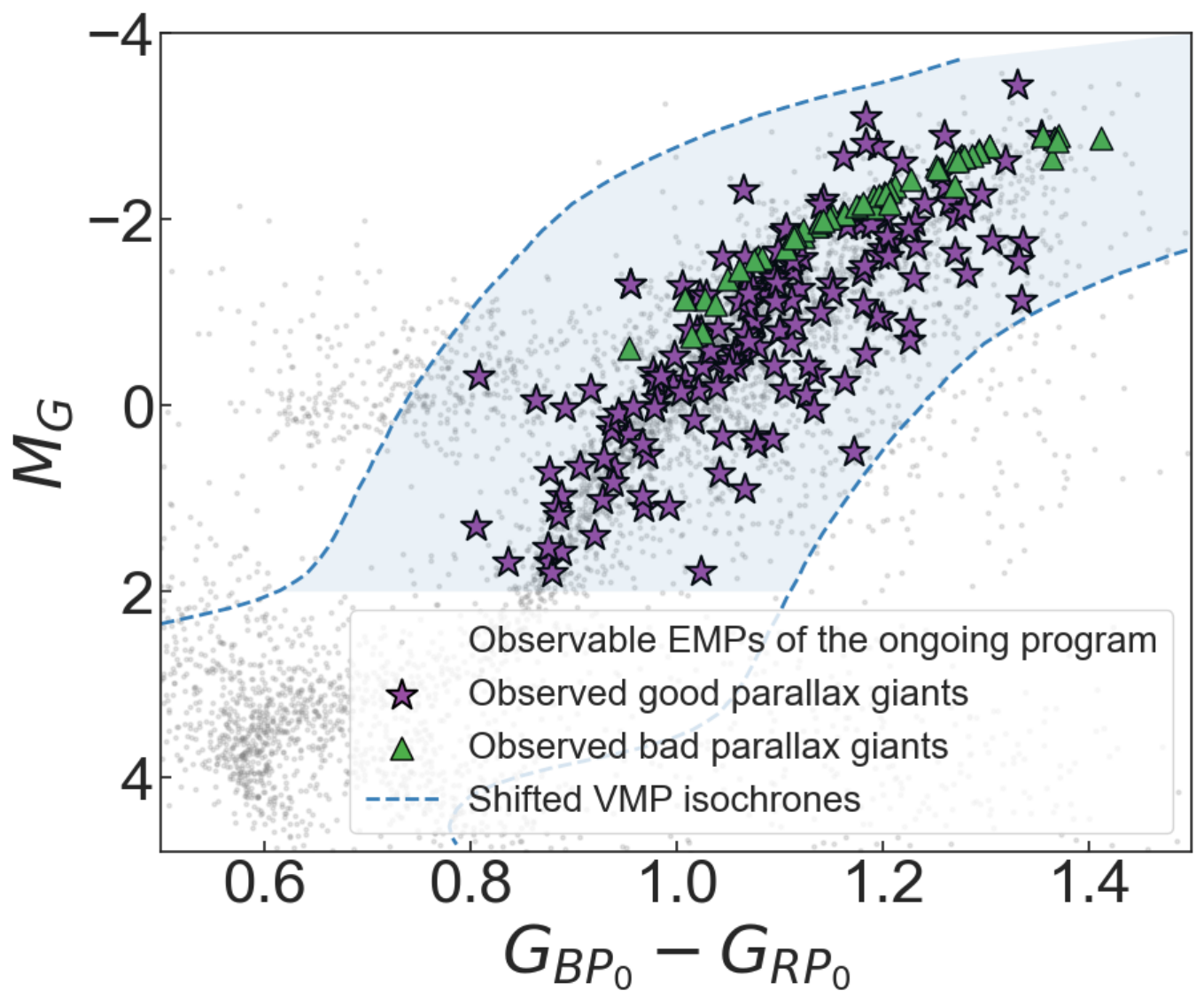}
    \caption{Colour absolute magnitude diagram (CaMD) of the INT observable targets that pass the quality cuts described in section \ref{2.2} in grey overplotted with the RGB stars that were observed and presented in this paper in purple and green. The purple star markers correspond to the good-quality parallax used to select the RGB stars and the green triangles correspond to the bad-quality parallax used to select the RGB stars for which the distances were derived using photometry and isochrone fitting. The blue dashed lines refer to an age 13 Gyr and metallicity [M/H] -2.2 PARSEC isochrone shifted by $\pm0.2$ mag in $BP_0-RP_0$ and $\pm0.75$ mag in $M_G$. The blue polygon is the absolute magnitude limited ($M_G<2.0$) target selection area.}
    \label{target-selection}%
\end{figure}

The creation of the Pristine-\textit{Gaia} synthetic catalogue is described in detail in \citetalias{2023martin}. We summarise the method here briefly. With the newest data release of \textit{Gaia} \citep[][DR3]{2022gaia}, the spectrophotometric BP/RP information \citep{2023deangelinew} was used to construct a comprehensive catalogue of synthetic \textit{CaHK} magnitudes, which emulate the narrow-band photometry employed in the Pristine survey. 
Several other recent works released a catalogue of metallicity estimates based on these Gaia XP spectra \citep{2023lucey,2023andrae,2023zhang}.
At the same time, the excellent \textit{Gaia} accuracy allows us to reprocess and recalibrate the entire Pristine \textit{CaHK} dataset comprising approximately 11,000 images obtained since 2015, resulting in an updated survey that now covers more than 6,500 square degrees.

The improved photometric catalogue exhibits enhanced accuracy, achieving a precision level of 13 millimagnitudes (mmag), a notable improvement from the estimated 40 mmag during the first year of data. 
As part of the updated model, the Pristine method for deriving the photometric metallicity of a star from \textit{CaHK} magnitudes and broadband magnitudes now relies solely on \textit{Gaia} broadband information (G, G$_{BP}$, G$_{RP}$), as opposed to SDSS previously.
Additionally, a more reliable iterative approach for addressing extinction correction has been implemented. 
This involves incorporating extinction correction on the \textit{Gaia} broadband magnitudes and \textit{CaHK} synthetic (or Pristine \textit{CaHK}) narrowband magnitudes by considering the star's photometric temperature and metallicity. 
To mitigate the impact of photometrically variable sources, which can introduce spurious metallicities, a variability model based on the photometric uncertainties of the 1.8 billion \textit{Gaia} sources was also implemented in the catalogue. 
Employing both the Pristine \textit{CaHK} magnitudes and the BP/RP based synthetic \textit{CaHK} magnitudes within the Pristine model, two catalogues of photometric metallicities for reliable stars were made public: the Pristine-\textit{Gaia} synthetic catalogue and the Pristine data release 1 (DR1) catalogue of photometric metallicities, encompassing stars common to both Pristine and the BP/RP catalogue of \textit{Gaia} DR3. 
The latter served as the first data release of the Pristine survey and provides deeper and better signal-to-noise (S/N) data for stars in common.

Both these catalogues facilitate the construction of reliable samples of metal-poor stars, with a particular emphasis on tracking V/EMP stars. 
The Pristine-\textit{Gaia} synthetic catalogue provides photometric metallicities across a vast portion of the sky, while the Pristine BP/RP catalogue, limited to the Pristine survey's footprint, offers notably high-quality metallicities, extending to significantly fainter stars. 

In this work, we use the Pristine-\textit{Gaia} synthetic catalogue of photometric metallicities to select V/EMP targets that have reliable distances and are on the red giant branch (hereafter RGB) of stellar evolution for a dedicated spectroscopic follow-up. We choose to follow-up red giants as they are intrinsically bright and probe large distances out into the Galactic halo, allowing us to study the chemodynamics of the Milky Way out to large distances and down to very low metallicities. For this target selection, we use the following quality cuts on the parent sample, most of which are recommended by \citetalias{2023martin}. Because some of the cuts recommended by \citetalias{2023martin} were introduced after this follow-up program commenced, these cuts are only implemented later in the manuscript as stricter quality cuts in section \ref{4.1}. The quality cuts initially used for target selection are defined as follows:

\begin{figure*}
    \centering
    \includegraphics[width=\textwidth]{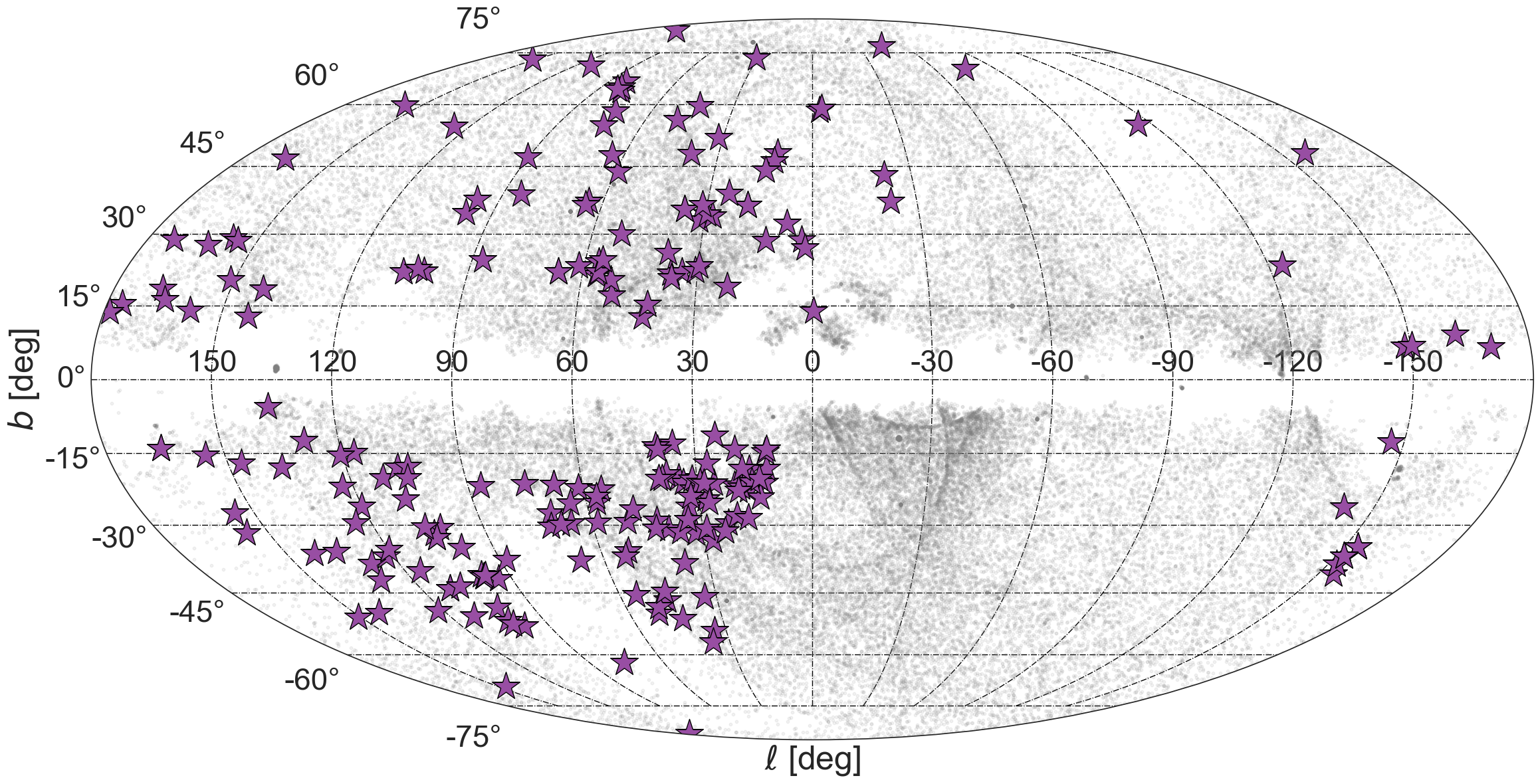}
    \caption{Mollweide projection of the Galactic coordinates for the V/EMP stars with [Fe/H]<-2.5 in the Pristine-\textit{Gaia} synthetic catalogue (in grey). The large-scale patterns visible in the map are mainly attributed to the scanning law of the \textit{Gaia} satellite. The purple stars represent the subset of stars observed and analysed in this spectroscopic follow-up.}
    \label{allsky-targets}%
\end{figure*}

\begin{itemize}
    \item The follow-up is done from the northern hemisphere. Declination greater than -30 degrees - which is the observability of the INT/IDS telescope facility used for the follow-up (\texttt{dec}>-30).
    \item Photometric metallicity lower than -2.5 dex (\texttt{FeH\_CaHKsyn}<-2.5). Note that this cut was based on the first generation of photometric metallicities using the Pristine survey model on \textit{Gaia} XP based \textit{CaHK} magnitudes. Over the course of this work, the photometric metallicities were updated as a result of improvement in the methods. Most of the selected stars still have a V/EMP metallicity inference, also in the published catalogue \citepalias{2023martin}. In the rest of the paper, we only ever list the final \citetalias{2023martin} metallicities.
    \item Percentages of Monte Carlo iterations used to determine [Fe/H] uncertainties inside the grid is greater than 80\% (\texttt{mcfrac\_CaHKsyn}>0.8)
    \item 84th percentile value of the probability distribution function (PDF) of the photometric metallicity is greater than -3.999 (\texttt{FeH\_CaHKsyn\_84th}>-3.999)
    \item   Photometric metallicity uncertainty less than 0.5 dex (0.5*(\texttt{FeH\_CaHKsyn\_84th} - \texttt{FeH\_CaHKsyn\_16th})<0.5 dex)
    \item Extinction on B-V magnitude is less than 0.5 (\texttt{E(B-V) }<0.5)
    \item Brightness cut on measured \textit{Gaia} G magnitude (\texttt{phot\_g\_mean\_mag<15.5}) to be able to observe with INT/IDS with short exposure times. 
    \item Photometric quality cut that is defined as {$C^*$}<{3$\sigma_{C^*}$}. Cstar is Gaia DR3 corrected flux excess, C$^*$, as defined in equation 6 of \citealt{2021riello} and Cstar\_1sigma is normalised standard deviation of C$^*$ for the G magnitude of this source, as defined in equation 18 of \citealt{2021riello} (abs(\texttt{Cstar})<\texttt{3*Cstar\_1sigma})
    \item Parallax cuts:
    \begin{itemize}
        \item Good parallax giants: \texttt{parallax\_over\_error>5}
        \item Bad parallax giants: Poorly constrainted small parallax (very likely a red giant) - \texttt{parallax\_over\_error}<5 and abs(\texttt{parallax})<0.2. 
    \end{itemize}
    \item No star within radius r$_{max}$ from the centre of globular clusters, with r$_{max}$ being the rough estimate of the size of a cluster defined by \cite{2021vasiliev}
    \item Shifted VMP old isochrone \citep[PARSEC isochrone\footnote{\href{http://stev.oapd.inaf.it/cgi-bin/cmd}{stev.oapd.inaf.it/cgi-bin/cmd}} of age 13 Gyr and metallicity -2.2 dex,][]{2012bressan,2020pastorelli} selection to remove horizontal branch stars and stars away from RGB (used to reliably select RGB stars as targets) - this selection is also limited to absolute magnitude of 2 mag and below. The magnitude-limited shifted isochrone selection is shown as the filled blue region in Figure \ref{target-selection}. 
    \item No overlap with literature high-resolution follow-up of VMP stars \citep{2018hansen,2018sakari,2018li,2022li} and the Pristine survey training sample that also includes Pristine survey's own spectroscopic follow-up programs as described in \citetalias{2023martin}. We end up having a handful of stars (14) that overlap with the SAGA database of metal-poor stars \citep{2008suda} (mostly from \citet{2022li} follow-up to avoid uncertainties due to many different follow-up methods) which we use to define the systematic uncertainty in the spectroscopic metallicity inferred in this work. 
\end{itemize}

All the stars that pass the above selection criteria (using good parallax subsample) with inverted parallax as an approximation for distance (=1/parallax) are shown in grey in Figure \ref{target-selection}. 
The reliable parallax stars that were spectroscopically followed-up are shown as purple star symbols and the ones that were selected based on small parallax with large errors are shown as green triangle symbols. 

Numerous studies have provided compelling evidence that relying solely on inverted parallax measurements to estimate distances can yield inaccurate results especially if the parallax measurements are imprecise ($\varpi$<0" and/or $\varpi/\sigma_{\varpi}$>0.2).
Incorporating additional priors and/or photometry has proven to enhance distance estimation \citep[e.g.,][]{2018bailer-jones,2021bailer-jones, 2018QUEIROZ,2022anders}.
To address these challenges, we adopted a Bayesian approach to infer distances for the stars with bad parallax in our sample. 
We employed a methodology, that enables the estimation of the PDF or posterior on the distance estimates (see \cite{2019sestito} for a full description of the method). 
In summary, the likelihood function is formulated as the product of Gaussian distributions from the parallax and Gaia photometry. Our prior incorporates a power-law stellar distribution for the halo that we believe to be more suited for VMP stars than the prior used in more general distance methods.
Additionally, we utilised VMP ([M/H] = -2.5) MESA/MIST isochrones\footnote{\href{https://waps.cfa.harvard.edu/MIST/index.html}{waps.cfa.harvard.edu/MIST}} \citep{2016dotter,2016choi} to account for the age and mass characteristics of VMP stars (11-13.8 Gyr, <1 M$_\odot$) and their distribution according to an IMF-based luminosity function in the colour absolute magnitude diagram (CaMD). 

This Bayesian framework, extensively employed in chemodynamical investigations of VMP stars \citep[e.g.][]{2019sestito,2020sestito,2023sestito,2020venn}, yields distance estimates with low uncertainties even when faced with significant parallax uncertainties. 
Typical distance uncertainties for these stars in the sample is as low as $\sim$8\%.
This is achieved because the MIST isochrones restrict the potential distance solutions for a star with a given colour to two possibilities: a dwarf or a giant solution. 
The parallax measurement typically favours one of the two solutions, or in the case of a very poor parallax measurement, assigns different probabilities to the two peaks. 
This approach eliminates the possibility of intermediate distance solutions. 
With this method all our targets that were selected based on poorly constrained small parallax ended up having a PDF that favoured a giant solution. 
These are shown as green star symbols along a rough isochrone-like positioning on the CaMD in Figure \ref{target-selection}. 
All the \textit{Gaia} broadband photometry in Figure \ref{target-selection} and other photometry used in target selection criteria are corrected for extinction based on the iterative method described in detail in \citetalias{2023martin} and available in their public catalogue. 
The on-sky distribution of the stars selected and observed with the telescope facility is shown as purple stars in Figure \ref{allsky-targets} while the grey stars in the background are all stars that satisfy the quality cuts described above (except the observability cut). 
The effect of \textit{Gaia}'s scanning pattern, which has resulted in certain regions being more frequently observed, is clearly evident in the distribution of grey points in Figure \ref{allsky-targets}. This pattern becomes particularly prominent as we push the limits of our observations to detect fainter stars, approaching \textit{Gaia} XP spectra's limiting apparent magnitude (for a more detailed visualisation of \textit{Gaia}'s scanning pattern in the analysis of sources with XP spectra, see \citealt{2023deangelinew})






\section{INT/IDS spectroscopic follow-up of V/EMP candidates}\label{3}

\begin{table*}
    \caption{Technical information of the facility used in this analysis}             
    \label{telescope}      
    \centering  
    \setlength{\tabcolsep}{0.5em}
    \begin{tabular}{ccccccccccc}     
        \hline\hline       
        Telescope & Instrument & Detector & Grating & Filter & Slit & Central $\lambda$ & Range & Dispersion & Resolving power ($\Delta\lambda$)\\ 
        \hline                        
        2.5m INT & IDS & Red+2 & R1200R & GG495 & 1.37" & 8500 $\AA$  & 7829.5 - 9163.7 $\AA$ & 0.51 $\AA$ px$^{-1}$ & 8019 (0.51 $\AA$) \\
        \hline\hline               
    \end{tabular}
\end{table*}

The spectroscopic follow-up presented in this paper stemmed from an ongoing long-term program at the \textit{Isaac Newton} Telescope (INT) in La Palma (PI: A. Viswanathan, semesters 22B, 23A, 23B, PI: E. Starkenburg, A. Arentsen, semester 22A). 
We followed-up 215 stars over the course of 17 nights (along with a few other member candidates from substructures in the Milky Way). The observations include the following grey to bright nights: June 17 to June 21 2022 (clear weather, seeing between 0.5 and 0.8), July 19 to July 21 2022 (clear weather, seeing $\sim$0.8), 8 September to 11 September 2022 (partly cloudy, seeing $\sim$0.9) and 30 December 2022 to 3 January 2023 (partly cloudy, seeing $\sim$1.4).
Our observing strategy was to go for very bright targets when the seeing is relatively bad, stay 40$^\circ$ away from the moon and away from the clouds. 
The results from semesters 23A and 23B will be published in a forthcoming paper.
We used the Intermediate Dispersion Spectrograph (IDS) equipped with the RED+2 CCD. 
Our configuration included the R1200R grating, a 1.37" slit width, and the GG495 order-sorting filter. 
This instrumental setup allowed us to achieve an effective spectral range of approximately 7850 to 9150 $\AA$ with a low to medium resolving power of $\sim$8000 (0.51 \AA) over 2 pixels at the detector, specifically at 8500 $\AA$, covering the calcium triplet region. 
This choice was made due to the follow-up of stellar streams (the original proposal that was changed to follow-up of EMPs in the later semesters with the use of a blue CCD).

 
The combination of the INT and the IDS instrument provided the necessary capabilities to achieve accurate metallicity information and radial velocities for the observed V/EMP candidates (see Table \ref{telescope} for further technical details).

\subsection{Data reduction}\label{3.1}

The spectra were processed and reduced using the Image Reduction and Analysis Facility \citep[IRAF,][]{1986tody} software package. 
Standard reduction techniques were applied, including bias sky subtraction, removal of telluric
lines, flat fielding, spectrum extraction, sky subtraction, wavelength calibration - using CuNe+CuAr
lamps and the ONEDSPEC package in IRAF, and heliocentric radial velocity correction. For wavelength calibration, we use a 10s arc per target and use the 10s and 30s arc from the beginning of the night for the calibration itself.
The fringing effect, which can sometimes be a mild issue for the RED+2 CCD on the INT\footnote{\url{https://www.ing.iac.es/astronomy/instruments/ids/ids_redplus2.html\#4.}}, was found to have an amplitude of 2\% or less for the redmost wavelengths below 9000 $\AA$. 
Since the wavelength range concerned by this study is covered by the calcium II triplet from approximately 8400 $\AA$ to 8750 $\AA$, no fringing correction was necessary.

\begin{figure*}
    \centering
    \includegraphics[width=\textwidth]{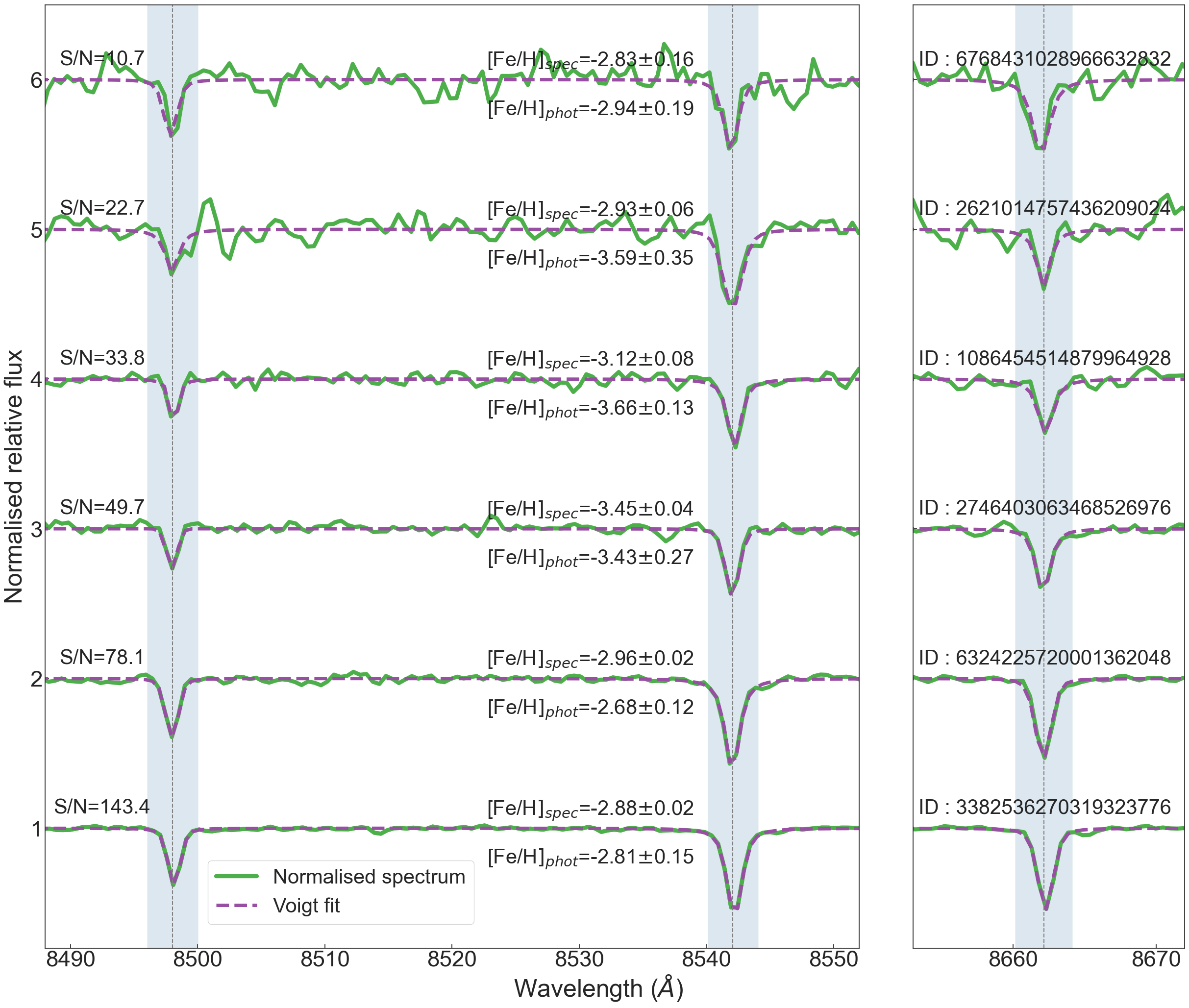}
    \caption{INT/IDS spectra of six V/EMP subsample stars selected using photometric metallicities from the Pristine-\textit{Gaia} synthetic catalogue, centred on the calcium triplet lines. The stars are presented and sorted based on their S/N. The normalised spectra are shown as solid green lines, while the fits, derived from our pipeline for Voigt line profiles detailed in section \ref{3.3}, are shown as dashed purple lines. Their S/N values, the photometric metallicities used to select them, and the spectroscopic metallicities derived are also given near the corresponding spectra. For clarity, the normalised flux of each star is shifted by 0.5 and the spectra between the second and the third calcium triplet line is cut off. The calcium triplet lines are highlighted in blue and the central line is indicated by grey dashed lines. We focused on the three calcium triplet lines and removed the core between the second and third calcium triplet lines only for clarity of the spectra presented in this figure and for no scientific purpose.}
    \label{spectra}%
\end{figure*}

Figure \ref{spectra} illustrates spectra of six examples with varying S/N between and 10 and 140. 
The S/N values for each spectra are shown above each of the spectra.
The S/N values calculated per pixel were computed in the calcium triplet region between line 2 (8542 \AA) and line 3 (8662 \AA). 
Due to the brightness range of the targets, (see left panel of Figure \ref{gmag-feh-distribution}) we have achieved a mean S/N of 34.2 with a minimum S/N of 9.7 with an average exposure time of 10 minutes (the exposure time varies per target based on the brightness of the target and seeing conditions).
We use a sophisticated MCMC pipeline to analyse the spectra to get metallicities (described in later subsections). This pipeline needs a S/N of $\geq$ 3 in the calcium triplet region to be able to statistically distinguish between the calcium triplet line with the smallest amplitude and noise \citep[see][]{2022longeard}. However, to achieve more robust results for our relatively bright stars, we aim for a S/N $\geq$ 15. The aimed S/N of 15 or above was achieved for $\sim$97\% of our targets while all our targets have a S/N above 3.

\subsection{Radial velocities}

To derive the radial velocities from calcium II triplet lines, we first cut the spectra within the wavelength range $\lambda$ = [8400, 8750] $\AA$, remove the cores of the calcium triplet (only to calculate the continuum) using \texttt{dispcor}, to find the continuum, and normalise the spectra using \texttt{onedspec.continuum} routine. We then run \texttt{fxcor} to obtain radial velocities through the cross-correlation of an observed standard star\footnote{Radial velocity standards are chosen from this table: \href{http://obswww.unige.ch/~udry/std/stdnew.dat}{obswww.unige.ch/~udry/std/stdnew.dat}} from the night with the observed V/EMP target. This is done by relative radial velocity between the observed target and the standard star and then subtracting the radial velocity of the standard star from the literature within \texttt{fxcor} module. 
All the routines used here are available in PyRAF and IRAF. 
At this stage of the analysis, the spectra has undergone several preprocessing steps including normalisation, radial velocity shift, removal of cosmic rays, subtraction of sky background, and correction for telluric absorption. 
Resulting spectra are shown in green with the flux of each spectrum shifted by 0.5 for better visualisation in Figure \ref{spectra}. 
Around 2 $\AA$ around each of the three calcium triplet regions are indicated within the blue filled regions with the central wavelength as black dashed lines. 

\subsection{Metallicities}\label{3.3}

To compute the (very small) additional radial velocity offset that may have been caused due to the difference in standard stars' radial velocity and equivalent widths, we use a modified version of the pipeline defined and used by \cite{2022longeard} and \citet{2024viswanathan}. 
It is noteworthy that this radial velocity correction is negligible (less than 3\%).
The first step in our analysis is to estimate the initial radial velocity of the star. 
We begin by creating a smoothed spectrum for each star using a Gaussian kernel with a width corresponding to 4 elements of resolution. 
This smoothing helps highlight the calcium triplet (CaT) lines. 
We model each calcium triplet (CaT) line using a Voigt profile and determine their positions by minimising the square difference between a simulated spectrum containing only the modelled CaT lines and the observed spectrum. 
The Voigt profiles consider parameters such as the amplitude of the Lorentzian component, the standard deviations of both the Lorentzian (half-width at half-maximum) and Gaussian components, and the Doppler shift in wavelength linked to the radial velocity offset based on the definition in \citet{MCLEAN1994125}.

To obtain an initial estimate of the radial velocity, we perform a cross-correlation between the simulated spectrum and the observed spectrum. 
This initial estimate is typically close to zero (mean of $\sim$1 km/s) due to previous alignment to the rest frame using a standard star. 
The derived radial velocity is then used as a starting point for further analysis and to improve the accuracy of the radial velocity measurement from the \texttt{fxcor} pipeline in IRAF. 
The mean statistical uncertainty in the inferred radial velocity is $\sim$6.9 km/s.
The radial velocity measured from the INT spectra show good agreement with the radial velocity from \textit{Gaia}'s Radial Velocity Spectrometer (RVS) derived radial velocity for a large majority of our sample agreeing within 25 km/s with \textit{Gaia} RVS radial velocity, with the exception of $\sim$6\% of the stars with more than a 2$\sigma$ difference between the two measurements. For 2 stars, this difference is as large as 5$\sigma$. Such a large difference in radial velocity measurements suggests the possibility of these stars being part of a binary system. 
However, Gaia flags and astrometric quality parameters such as Renormalised unit weight error, \texttt{RUWE}, Total amplitude in the radial velocity time series after outlier removal, \texttt{RVamp}, and Radial velocity renormalised goodness of fit, \texttt{RVgof} do not suggest a binary origin. Some of these parameters are not available for all the stars but only for the brighter subsample (G<12). Due to this ambiguity, we refrain from using these members (2$\sigma$ and above, 14 stars) in the dynamical analysis of the paper. However, the 1D spectra and inferred paramaters are released for all our stars including these radial velocity mismatched stars. It is noteworthy that we do not find any systematic offset between the two radial velocities. We use the \textit{Gaia} RVS radial velocity (where available) for the dynamical analysis of this paper with 8 stars added using INT radial velocities for which \textit{Gaia} radial velocities are not available.

Next, we employ a Monte Carlo Markov Chain (MCMC) algorithm with a million iterations per spectrum to fit the observed spectra. 
Using the simulated spectrum that includes only the CaT lines shifted according to the initial radial velocity guess and the initial amplitude guess (which is set to start lower for EMP stars than normal metal-rich population in the Galaxy), we derive the final radial velocity and equivalent widths (EWs) of the lines. 
The step size defined to explore the parameter space to achieve the optimal acceptance ratio is set based on the S/N of the spectrum.
The fitting process involves simultaneously optimising the central wavelengths, normalised fluxes, and standard deviations of each line. 
Constraints are applied to ensure the relative depths and widths of the lines agree with respect to each other. 
For example, the second CaT line is constrained to be deeper than the third one which in turn is set to be deeper than the first one. 
We also ensure that the first line is not narrower than the other two lines, and the third line is narrower than the second line. 
These constraints help maintain the expected order and shape of the CaT lines during the analysis. 
The MCMC analysis is performed for each star, and the parameters that maximise the likelihood are selected as the best-fit values. 
The central wavelengths (and therefore the radial velocities, v), the normalised fluxes of each line a1, a2, and a3 as well as their standard deviations are fitted by minimising the likelihood that is defined as:

\begin{equation}
    \mathcal{L}_i = \frac{1}{\sigma_i\sqrt{2\pi}}\exp{\left(-0.5\frac{(s_{obs,i}-s_{sim,i})^2}{\sigma_i^2}\right)}
    \label{eq:likelihood}
\end{equation}

where $\sigma_i$ is the flux uncertainty on star i, $s_{obs,i}$ is the observed spectrum and $s_{sim,i}$ is the simulated spectrum of star i, all of which are vectors. 
To calculate the EWs, we integrate each CaT line within a 15 $\AA$ window centred on the line. 
The integration is performed on the best-fit simulated spectrum, and the resulting integrated values represent the EWs for each line.
The best-fit Voigt profile for each of the six example stars of varying S/N is shown as purple dashed lines in Figure \ref{spectra}. 

\begin{figure}
    \includegraphics[width=\columnwidth]{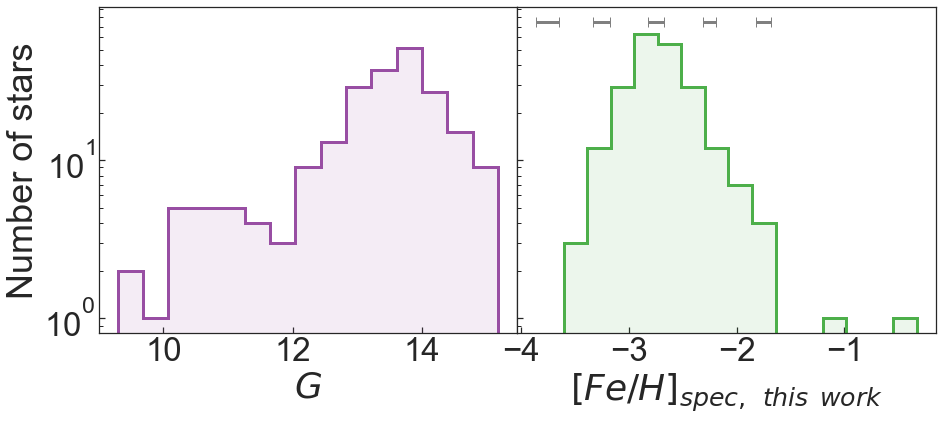}
    \caption{Distribution of \textit{Gaia} G magnitudes and the derived metallicities from the Ca II triplet region for the full follow-up spectroscopic sample of ~220 Pristine-\textit{Gaia} synthetic V/EMP stars. Median uncertainties in bins of 0.5 dex in spectroscopic [Fe/H] are shown on the right.}
    \label{gmag-feh-distribution}%
\end{figure}

To convert the equivalent widths (EWs) of the calcium triplet lines into metallicity measurements, we employ the calibration provided by \cite{2013carrera} for Voigt profile fits. 
This calibration is known to reliably estimate metallicities down to a value of -4.0.
The inputs for this conversion include magnitudes, calcium triplet equivalent widths, and distances (inverted parallax) or height above or below the horizontal branch. Magnitudes adjusted for extinction are taken from the input photometric metallicity catalogues. 
Converting from \textit{Gaia} G magnitude to Johnson-Cousins V or I magnitudes is achieved through the conversion prescribed by \citet{2021riello}.
To determine the uncertainties associated with the metallicity measurements, we utilise a Monte Carlo procedure. 
In each iteration, we randomly draw values of the EWs from their PDF. 
We then calculate the spectroscopic metallicity for each iteration, taking into account the individual uncertainties in photometry and distances.
Additionally, we incorporate the uncertainties associated with the calibration relation itself by considering the proposed uncertainties on the coefficients stated by \cite{2013carrera}. 
By performing this Monte Carlo process, we construct a PDF that captures the uncertainty in the metallicity determination for each star, considering all the relevant parameters involved in the spectroscopic metallicity derivation. 
The standard deviation (using a Gaussian approximation) of this distribution is taken as the uncertainty on the metallicities. 
The calibration for metallicities derived from the CaT equivalent width is grounded by \citet{2013carrera} in empirical observations of 55 metal-poor field stars. These stars were examined at high-resolution, R>20,000, measuring Fe I and Fe II spectral lines, which are less influenced by non-local thermodynamic equilibrium (NLTE) effects compared to the Ca II triplet lines. Consequently, the metallicities estimated through this calibration closely align with NLTE equivalents. 

\begin{figure*}
    \centering
    \includegraphics[width=\textwidth]{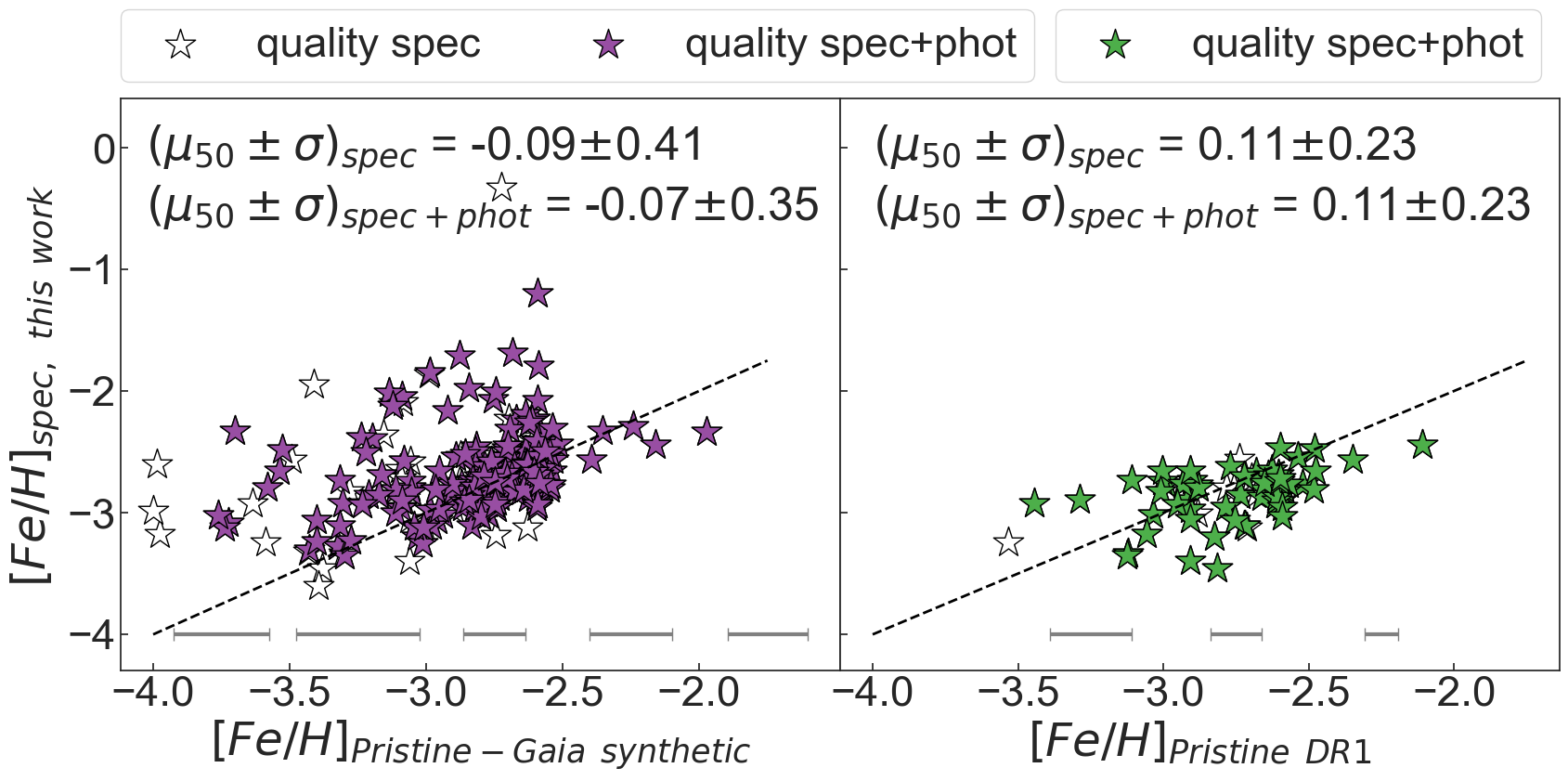}
    \caption{Comparison of photometric and spectroscopic metallicities for the full followed-up sample of stars. The left panel shows the photometric metallicities obtained using the Pristine-\textit{Gaia} synthetic (XP spectra) vs the spectroscopic metallicities derived from the calcium triplet equivalent widths for the entire sample. A small subset of the sample also has photometric metallicities derived from higher S/N Pristine \textit{CaHK} magnitudes, which are shown in the right panel vs the spectroscopic follow-up metallicities. Median uncertainties in bins of 0.5 dex in photometric [Fe/H] is shown in both panels. The uncertainties increase with decreasing metallicities in general, but the first and last bins are affected by low number statistics and do not follow this trend.}
    \label{feh-comparison}%
\end{figure*}

The brightness range probed by the sample and the spectroscopic metallicities obtained from this analysis are shown as 1D histograms in Figure \ref{gmag-feh-distribution}.
Four stars that are either too hot (spectra with no distinguishable calcium triplet lines indicating blue horizontal branch contamination) or have cosmic ray on the calcium triplet lines are removed from the analysis. The spectra for these stars and inferred parameters are not released with this work due to bad quality of the spectra. 
Futhermore, we also remove stars that have an absolute magnitude in Johnson-Cousins I-band outside the range, $-4<I<1$, within which the established EW to metallicity relation works reliably. 
The spectra for these stars and inferred parameters are not released with this work due to the fact that the conversion is not robust in this regime. 
Stars that pass all these quality criteria in spectra and magnitudes are referred to as 'quality spec' in Figure \ref{feh-comparison}. 
The median uncertainties on bins of 0.5 dex in metallicities is shown at the top of the histogram in the right panel of Figure \ref{gmag-feh-distribution}. 
We recommend to add to the reported measurement uncertainty (only $\sim$0.08 in the median), a systematic uncertainty of 0.15 dex derived from comparison with high-resolution analyses of VMP stars from the SAGA database. 
This is based on the 14 stars in overlap between our follow-up and the SAGA database. 
The remaining $\sim$200 stars are newly discovered V/EMP stars from this work, ideal for high-resolution follow-up. The on-sky distribution of the stars colour-coded by their inferred metallicities is shown in Appendix \ref{a}. 
We observed 225 stars as the first part of this spectroscopic campaign described in this work, and remove 4 stars due to being too hot (2 of them) or having cosmic rays on the CaT lines (2 of them), and 6 stars that have an I-band magnitude outside the range within which the EW to metallicity relation works reliably. This leaves us with a total of 215 stars. We release the 1D spectra and chemical and dynamical parameters inferred for these 215 stars observed and analysed in this work (including the 14 stars which has a radial velocity mismatch higher than 25 km/s with Gaia radial velocities as we have enough evidence to trust their analysis and inferred parameters). As part of this follow-up campaign, we have also observed more than 300 stars using the blue CCD (for which we will be able to infer carbon and $\alpha$ abundances along with metallicities), which will be analysed and released in an upcoming publication.

\section{Results}\label{4}

In this section, we present the findings of our spectroscopic analysis, which encompasses both the dynamical properties and the metallicity measurements of the observed stars.

\begin{table*}\label{pgs-success}
    \caption{Evaluation of the quality of the photometric metallicities from the Pristine-\textit{Gaia} synthetic catalogue.}   
    \label{comp}      
    \centering  
    \setlength{\tabcolsep}{0.5em}
    \begin{tabular}{llll}     
        \hline\hline       
        &Observed with recommended quality cuts&Stricter quality cuts&[Fe/H]$_{phot}$ < --2.5\\ 
        \hline 
        \\
        Total observed & 215 & 173 & 168\\
        $\text{[Fe/H]}_{phot}$ < --2.5 & 210 & 168 & 168 \\
        $\text{[Fe/H]}_{phot}$ < --3.0 & 60 & 38 & 38\\
        \hline
        $\text{[Fe/H]}_{spec}$ < --2.5 & 163 & 130 & 129\\
        $\text{[Fe/H]}_{spec}$ < --3.0 & 32 & 21 & 21 \\
        \hline
        $\text{[Fe/H]}_{spec}$ > --2.0 (outliers) & 9 (4\%) & 6 (3\%) & -\\
        success [Fe/H] < --2.0 & 206 (96\%) & 166 (97\%) & - \\ 
        success [Fe/H] < --2.5 & 162 (77\%) & 129 (77\%) & - \\
        success [Fe/H] < --3.0 & 23 (38\%) & 14 (37\%) & - \\
        \hline\hline               
    \end{tabular}\\
    \tablefoot{The number of stars with photometric metallicities from the Pristine-\textit{Gaia} synthetic catalogue below --2.5 and --3.0, the number of stars confirmed spectroscopically below those metallicity thresholds, and the corresponding success rates are provided for three distinct datasets: the entire spectroscopic sample after basic quality cuts, the sample post-application of the stricter selection criteria to increase the 1:1 agreement in photometric and spectroscopic metallicities, and the subset of stars with photometric metallicities below --2.5.}
\end{table*}

\subsection{Performance of the Pristine-\textit{Gaia} synthetic pre-selection}\label{4.1}

In the subsequent discussion of the results, we employ the terms "Pristine metallicities" and "photometric metallicities" interchangeably to denote the metallicity values obtained from the narrow-band synthetic \textit{CaHK} derived from \textit{Gaia} XP spectra combined with \textit{Gaia} broadband BP-RP data. Conversely, we use the terms 'CaT metallicities' and 'spectroscopic metallicities' to refer to the metallicities derived from the analysis of spectra acquired using the INT/IDS facility in the calcium triplet region. Figure \ref{feh-comparison} shows the one-on-one comparison between the spectroscopic metallicities of the V/EMP stars observed and analysed in this work versus their photometric metallicities from the Pristine-\textit{Gaia} synthetic catalogue in the left panel and a subsample (69) of stars that have photometric metallicities from the Pristine DR1 bright catalogue (higher S/N \textit{CaHK} narrowband values) in the right panel. 
The 1:1 line and median uncertainties in photometric metallicities in bins of metallicities are shown in both panels. 
The difference between photometric and spectroscopic metallicities have a median offset of -0.08 dex and 1$\sigma$ deviation of 0.41 dex for the Pristine-\textit{Gaia} synthetic catalogue and a median offset of +0.11 dex and 1$\sigma$ deviation of 0.23 dex for the Pristine survey data release 1 catalogue. 
The 1$\sigma$ deviation is reduced by about half with the higher S/N \textit{CaHK} narrowband measurements from the Pristine survey DR1 data. 
As mentioned in section \ref{2.2}, some of the quality cuts defined by \citetalias{2023martin} came into effect after the start of our follow-up program. 
We investigate how these cuts and some other stricter cuts can improve the one-to-one agreement of photometric and spectroscopic metallicities.
For the stricter quality cuts, we implement the probability of photometric variability to be less than 30\% (\texttt{Pvar} < 0.3, which was a quality cut established while the follow-up program was already ongoing), stricter photometric quality cut (abs(\texttt{Cstar} < \texttt{Cstar\_1sigma})), astrometric quality cut (Renormalised Unit Weight Error \texttt{RUWE} < 1.4, only three stars observed have ruwe greater than 1.4 and have been removed in the dynamical analysis section of this paper because they can be non-single or otherwise problematic astrometric solutions), photometric metallicity uncertainty less than 0.3 dex (0.5*(\texttt{FeH\_CaHKsyn\_84th} - \texttt{FeH\_CaHKsyn\_16th}) < 0.3 dex), percentage of Monte Carlo iterations for [Fe/H] uncertainties equal to a 100\% (\texttt{mcfrac\_CaHKsyn} == 1.0), 16th percentile value of photometric metallicity greater than -3.999 (\texttt{FeH\_CaHKsyn\_16th}>-3.999), and CASU photometric data reduction flag (\texttt{flag} = -1, denoting very likely point-sources - only for sources with Pristine DR1 measurements).
Stars that pass these extra quality cuts are referred to as "quality spec+phot" in Figure \ref{feh-comparison}.
The median offset and 1$\sigma$ deviation in the difference between spectroscopic and photometric metallicities reduces from -0.09 dex to -0.07 and 0.41 dex to 0.35 dex respectively with the stricter quality cuts in the input Pristine-\textit{Gaia} synthetic catalogues, at the expense of losing 20\% of the VMP stars followed-up.  
The 1$\sigma$ deviation between photometric and spectroscopic metallicities are lower ($\sim$0.23 dex) for Pristine DR1 photometric metallicities due to their high S/N \textit{CaHK} measurements. It is noteworthy that regardless of the stricter quality cut, the agreement is quite good between the spectroscopic and photometric metallicities, and while the stricter quality cuts remove few outliers, they also remove few interesting VMP stars. For this reason, our conclusion is that the stricter quality cuts do not improve the selection

Table \ref{comp} shows the number of stars present in various V/EMP photometric and spectroscopic metallicity bins for three sub-samples from this spectroscopic follow-up program: A total sample of 215 stars observed with this program selected based on the recommended quality cuts and other cuts mentioned in subsection \ref{2.2}, stricter quality cuts to improve the 1:1 agreement between spectroscopic and photometric metallicities as described above, and all stars with photometric metallicity estimates below --2.5 dex (a subsample 168 stars). 
Firstly, the table outlines the number of stars within each sample that were estimated by the Pristine-\textit{Gaia} synthetic photometric metallicities to have [Fe/H] values equal to or less than -2.5 or -3.0, respectively. It also presents analogous counts based on spectroscopic [Fe/H] values and the corresponding success rates. Success rates denote the fraction of stars anticipated to have [Fe/H] values below a specified threshold by photometric metallicities, which were indeed observed to possess spectroscopic [Fe/H] values below that threshold. Outliers are stars with spectroscopic [Fe/H]>--2.0. 
A similar exercise was performed by \citet{2017youakim} and \citet{2019aguado} using the medium spectroscopic follow-up of CFHT-based Pristine survey photometric metallicities. 
Comparing the success rates, we find that the success rates at finding stars with [Fe/H] < -2.5 and [Fe/H] < -3.0 has increased from 56\% and 23\% to 77\% and 38\% respectively and the outliers ([Fe/H] > -2.0) have gone down from 12\% to 3\%, with no catastrophic outliers ([Fe/H] > -1).
We refrain from making estimates for the Pristine DR1 catalogue due to limitations in the number of stars but these numbers are expected to be higher due to the higher S/N \textit{CaHK} measurements and deeper magnitude ranges probed by the Pristine survey. 
The improvement in success rates are mainly due to the well-defined selection criteria weeding out catastrophic outliers due to photometry, astrometry and limitations in the photometric metallicity model. 
Given these new and improved success rates, we predict that the photometric metallicities from the Pristine survey and the Pristine-\textit{Gaia} synthetic catalogues will bring more than 10,000-20,000 homogenously analysed EMP stars from the WEAVE low-resolution (R$\sim$5000) follow-up of Pristine EMP stars \citep{2024weave} and allow for unprecedented statistical analyses of the metal-poor Galaxy. 

\begin{figure*}
    \centering
    \includegraphics[width=0.55\textwidth]{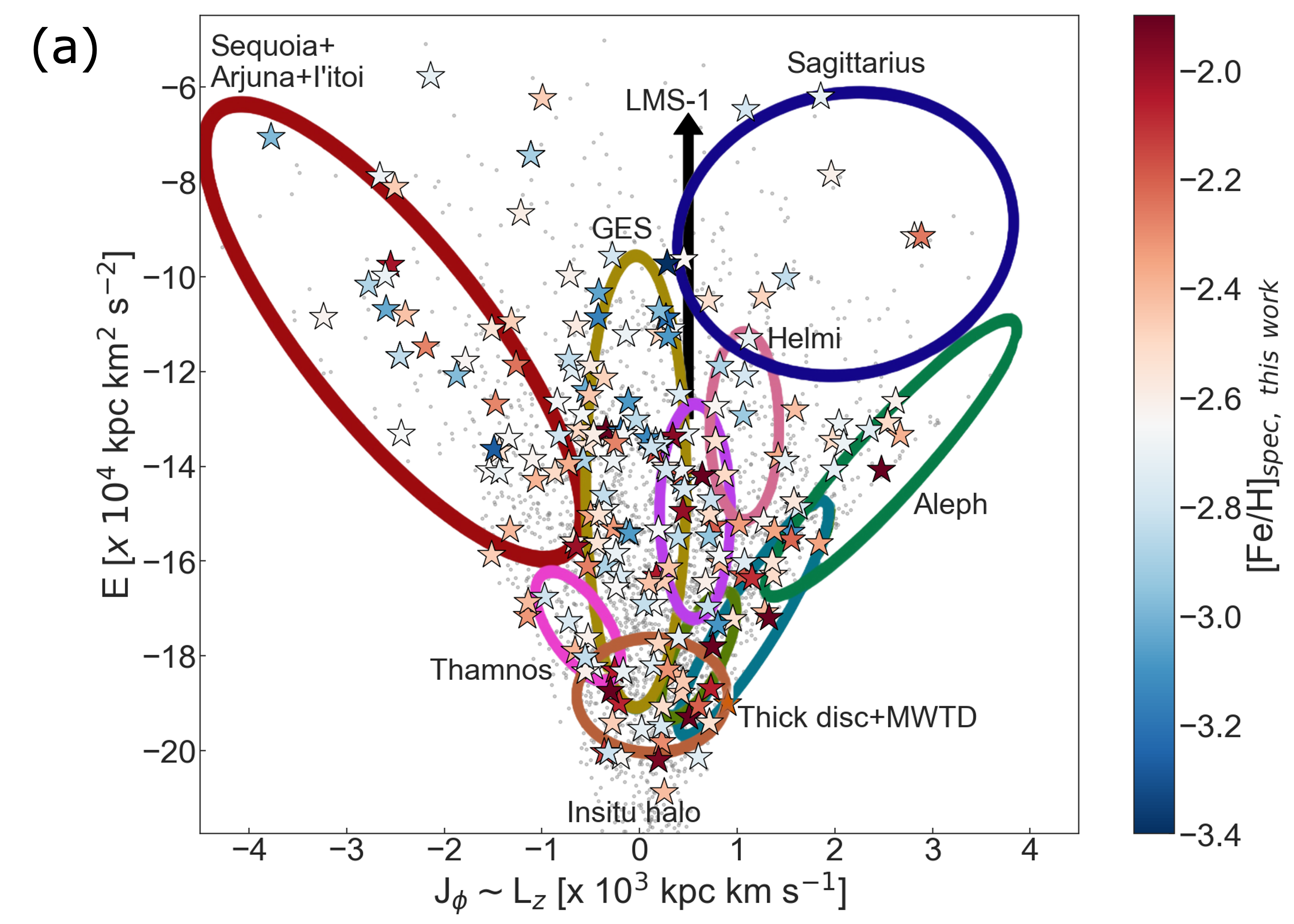}\\

    \includegraphics[width=0.95\textwidth]{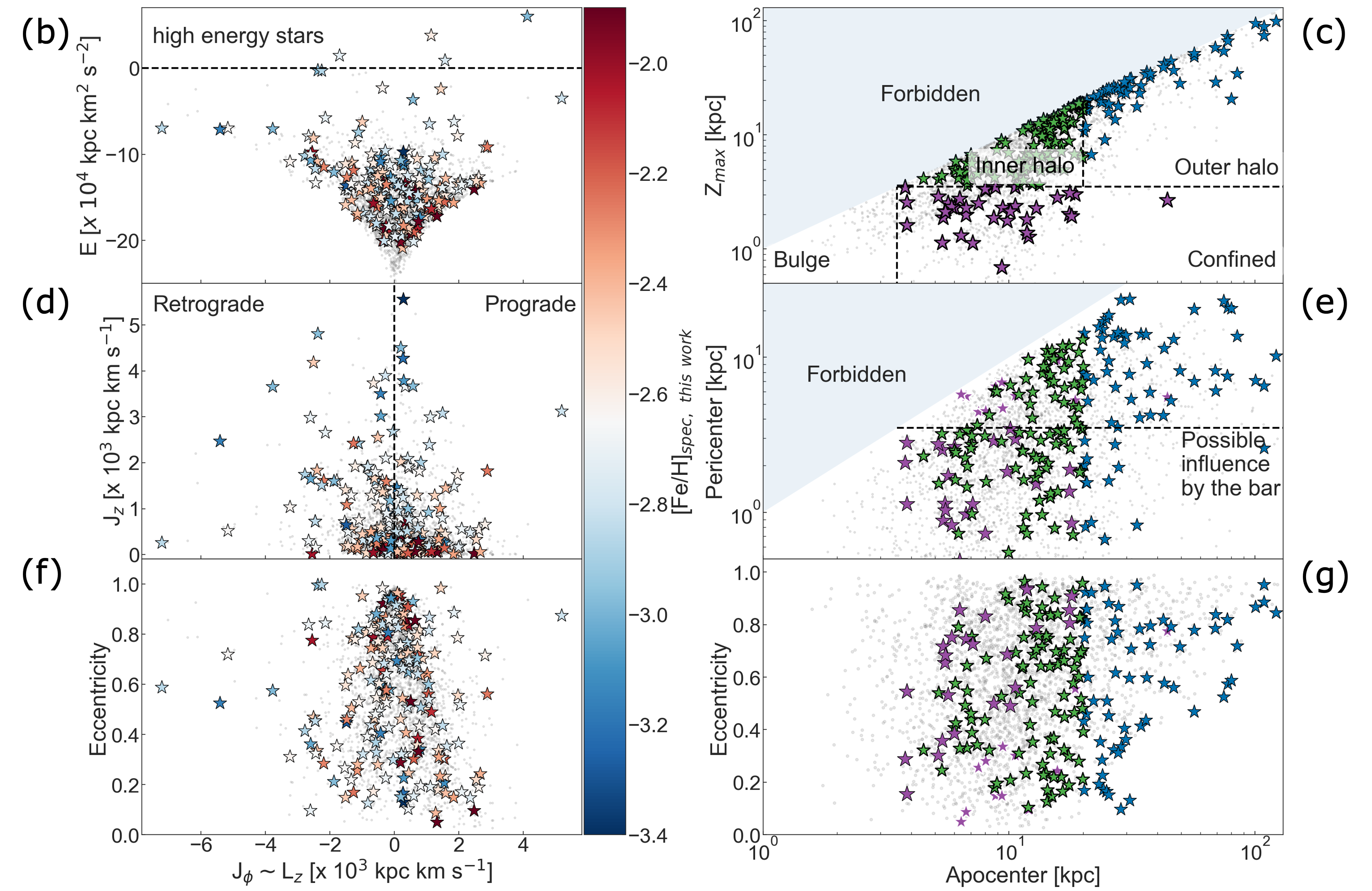}

    \caption{Chemokinematics of the V/EMP stars from this work. (Panel a) Energy E vs angular momentum in z direction L$_z$. The substructure regions from \citet{2020naidu} identified using RGB stars in the H3 survey are shown as coloured ellipses to associate V/EMP stars from this work. (Panel b) Same as top centre, but without the substructure ellipses. Stars with energy greater than 0 are high-energy and/or unbound stars in the Milky Way potential used. (Panel d) Rotational component of the action J$_\phi$ vs vertical action J$_z$. (Panel f) Rotational action J$_\phi$ vs eccentricity. 
    (Panel c) The INT sample is divided into three dynamical groups according to their apocentre and Z$_{max}$. The grey shaded area denotes the forbidden region in which Z$_{max}$ > apocentre. The vertical and horizontal lines separate the groups. (Panel e) Groups are shown in apocentre vs pericentre view. The grey shaded area denotes the forbidden region in which pericentre > apocentre. (Panel g) Groups in apocentre vs eccentricity view. All EMP stars observable by the spectroscopic follow-up program are shown in grey.}
    \label{kinematics}%
\end{figure*}

\subsection{Chemokinematics of the V/EMP stars}\label{4.2}

To calculate the orbital properties of V/EMP stars from our follow-up program, we use right ascension, and declination in J2016 format, proper motion in right ascension and declination from \textit{Gaia} DR3 data, with distances based on parallax (from \textit{Gaia} DR3) or photometry as described in section \ref{2.2} and radial velocities from \textit{Gaia} RVS from DR3 when available and INT spectra based radial velocities otherwise (for 6 stars). We adjust the stars for solar motion, considering (U$_\odot$, V$_\odot$, W$_\odot$) = (11.1, 12.24, 7.25) km/s according to \citet{2010schonrich}, and for the motion of the local standard of rest (LSR) using v$_{LSR}$ = 232.8 km/s as per \citet{2017mcmillan}. Both the Galactocentric cartesian and cylindrical positions and velocities of the stars are computed assuming R$_\odot$ = 8.2 kpc \citep{2017mcmillan} and z$_\odot$ = 0.014 kpc \citep{1997binney}. Our coordinate system is oriented such that x points towards the Galactic centre, y aligns with the direction of motion of the disc, and positive (negative) z signifies the height above (below) the disc. Angular momenta, L$_z$ and L$_\bot$ = $\sqrt{\text{L}_x^2 + \text{L}_y^2}$, are calculated for the stars with the sign of L$_z$ inverted to ensure it being positive for prograde orbits. Energy, E, is determined using AGAMA \citep{2019vasiliev} and the Milky Way potential from \citet{2017mcmillan}. This potential, which is axisymmetric, encompasses a stellar thin and thick disc, an HI gas disc, a molecular gas disc, a bulge, and an NFW halo, which defaults to a spherical shape. 
The integration time is set to 1 Gyr. 
The output orbital parameters are the Galactocentric Cartesian coordinates (x, y, z), the maximum distance from the Milky Way plane Z$_{max}$, the apocentric and pericentric distances, the eccentricity, the energy E, and the spherical actions coordinates (J$_\phi$, J$_r$ , J$_z$). 

In panel b of Figure \ref{kinematics}, we see the total energy (E) versus the z-component of angular momentum (L$_z$). 
We see four stars (all below -2.5 in [Fe/H]) that are unbound in our chosen gravitational potential of the Milky Way, along with 7 stars that are not associated with any known substructure region and have high apocentres. 
These unbound stars have a velocity of about 700 km/s at a distance of about 15 kpc which is too small to classify them as hyper velocity stars.
From the literature, we see that most of the hyper velocity stars are coming from the Galactic disc. These four high-energy stars also have confined orbits (Z$_{max}$ < 3.5 kpc).
However, given that their proper motions and velocities are small, it is safe to say that these stars are most likely unbound due to the measurement uncertainties.
In the future, after a careful target selection of high proper motion or high-velocity stars, a large sample of stars in high-energy orbits can be used to infer the escape velocity of the Milky Way. We note that the high-energy stars are not shown in the left panels, due to their very high apocentre and Z$_{max}$.

It is anticipated that stars originating from the same physical structure will exhibit clustering in integrals of motion, energy and angular momentum (especially L$_z$) within an axisymmetric potential, despite being widely dispersed in configuration space \citep{2000helmi}. 
Since the second data release DR2 of the \textit{Gaia} mission, phase mixed streams have been identified in kinematically selected halo stars. These structures, including the Gaia-Enceladus-Sausage, Sequoia, Thamnos, Helmi streams, LMS-1/Wukong, Heracles/Kraken/Koala, Cetus-Palca as discussed in \citet{2019koppelman,2019myeong,2020yuan,2020naidu,2022lovdal,2022ruizlara,2022malhan,2022thomas,2023horta,2023dodd}, are believed to have an accreted origin primarily based on their inferred orbits, often supported further by chemical abundances where available.
Nevertheless, their chemical composition, essential for determining the types of galaxies from which they originated, remains inadequately characterised, particularly at the lowest metallicities.
Therefore, we associate our V/EMP stars with known substructures with ellipses as detected using red giant stars from the H3 survey \citep{2020naidu} in the integral-of-motion (IOM) space (defined using the substructures in \citet{2017mcmillan} potential in Figure 23 from the same work). This is shown in panel a of Figure \ref{kinematics}).  
We associate stars with a particular accretion event based on the selection criteria presented in \citet{2020naidu}. Most significantly we can associate stars with the Sequoia+Arjuna+I'itoi group (23 stars with a huge spread in distance between 2-32 kpc, see also the small clump of the 3 VMP and EMP stars clustered tightly in this region and panel d), GES (total of 81 stars with $\sim$97\% within 20 kpc, note the high-eccentricity stars clumped around L$_z\sim$ 0 in the left bottom panel) and the metal-poor LMS-1/Wukong (14 stars with medium eccentricity, the slightly prograde clump of stars in panel f). 
We associate a handful of EMP stars with the Helmi streams (6), Thamnos (11), and Sagittarius (4), with both Thamnos and the Helmi streams populating distances less than 10 kpc while Sagittarius stars all lie above 7 kpc. 
There are also a few stars in the thick disc and the \textit{in situ} halo regions. 
We also find small clumps of stars with very similar metallicities that could originate from low-mass merger events or globular cluster disruptions.
These groups are indicated with a group number in Table \ref{catalogue}, which will be released electronically with this work.
Due to the low number statistics and complex selection function, we refrain from defining any new substructures from this work. 
In the future, a more complete and large sample of V/EMP stars with dynamics and chemistry will allow us to also identify the smallest and earliest galaxies that merged into the Milky Way \citep{2020yuan}. 

In Figure \ref{kinematics}, the three panels on the right (panels c, e, and g) illustrate the pericentric distance, eccentricity, and maximum height from the plane relative to the apocentric distance. Meanwhile, the right-hand three panels depict the energy versus the rotational component of the action (top), rotational versus vertical action (middle) and rotational action versus eccentricity (bottom).
In panel d, we see stars separated into prograde and retrograde orbits. We clearly see that our sample has more prograde than retrograde stars.
The sample is classified into four distinct groups based on their Z$_{max}$ and apocentre, akin to the categorisation outlined in \citet{2023sestito}. These categories are described as follows:

\begin{itemize}
    \item Bulge group: Stars within this group remain confined within a sphere with a radius of 3.5 kpc from the Galactic centre, i.e., apocentre < 3.5 kpc. No spectroscopic follow-up has been conducted in this region.
    \item Confined group (purple stars): Stars within this group exhibit Z$_{max}$ < 3.5 kpc and apocentre > 3.5 kpc. They are primarily restricted near the Milky Way disc, and this group comprises 41 (20\%) stars that have undergone spectroscopic follow-up and presented in this work
    \item Inner halo group (green stars): Comprising 101 (46\%) stars, this group consists of stars with Z$_{max}$ > 3.5 kpc and apocentre < 20 kpc.
    \item Outer halo group (blue stars): This group comprises 73 (34\%) stars characterised by Z$_{max}$ > 3.5 kpc and apocentre > 20 kpc.
\end{itemize}

As depicted in panel g of Figure \ref{kinematics}, all stars, regardless of their group, show a range of eccentricities. It is noteworthy that stars in the confined group do not necessarily exhibit planar orbits. If we define planar orbits as stars with a ratio of Z$_{max}$/apocentre < 0.2, these constitute half of the stars in our confined group. Of these stars, about 80\% are in prograde disc-like orbits. These findings are discussed in detail in the next section. 

In panel e of Figure \ref{kinematics}, we see that about 71\% of the stars that are confined have small pericentre comparable to the length of the Galactic bar \citep[R$_b\sim$ 3.5 kpc,][]{2015wegg,2023lucey}. 
\cite{2023dillamore} suggest that a possible co-rotation resonance of a centrally concentrated halo population with the rotating bar could explain why we see planar prograde stars in the V/EMP end in the solar neighbourhood. 
From our sample of confined stars (of which about half of them are prograde planar), indeed the bar can play a role in the orbit of this population. 
However, we do see a population of confined stars (small purple symbols in panels e and g of Figure \ref{kinematics}) that have higher pericentres and low eccentricity, more like disc orbits. The on-sky distribution of the stars colour-coded by their categories (confined, inner halo, outer halo) are shown in Appendix \ref{a}. Other dynamical spaces such as position distribution, velocity distribution, and IOM space for these categories of stars are shown in Appendix \ref{b}.

\section{Discussion}\label{5}

In this section, we discuss in detail the implications of planar metal-poor stars in the sample and present and discuss new members of the most metal-poor stellar stream C-19. 

\subsection{Planar metal-poor stars}\label{5.1}

The oldest and most metal-poor stars are usually found in the bulge and halo because they are relics from the era of the smallest, earliest galaxies that merged into the Milky Way halo. However, there have been recent discoveries of very, extremely and ultra metal-poor stars in prograde, planar, and disc-like orbits confined to the plane \citep[z < 3 kpc,][]{2019sestito}. The mechanisms for producing such stars beyond just the halo has been highly debated. Some of the proposed scenarios are as follows: (i) minor mergers that sink onto the plane before disrupting due to dynamical friction \citep{2003abadi,2019sestito}, (ii) stars from the building blocks of the proto-Milky Way or filaments creating a halo that is slowly rotating prograde \citep{2019sestito,2020sestito,2022belokurov,2023belokurov,2022rix,2023zhang}, (iii) the \textit{in situ} formation of this component of the disc at early times \citep{2019sestito,2020sestito,2020dimatteo,2021alvar,2024alvar,2024nepal}. Regarding the feasibility of the latter scenario, in the NIHAO-UHD and FIRE simulations these stars are formed within 2-3 Gyr from the Big Bang, while the disc appears later \citep{2021sestito,2021santistevan}. 
However, disc formation in simulations cannot be evidence enough to rule out this scenario.
(iv) A possible co-rotation resonance of a halo population with the bar \citep{2023dillamore}. 
\citet{2023yuan} illustrated that as a result of a rapidly decelerating bar, certain bulge stars acquire rotational motion by becoming ensnared in co-rotating regions and subsequently migrating outwards along prograde planar orbits. However, the proportion of stars influenced by this mechanism is insufficient to explain all of the metal-poor rotators identified. Consequently, their origin remains ambiguous. 

\citet{2019sestito} conducted an analysis of the kinematics and dynamics of all ultra metal-poor (UMP, [Fe/H] < -4.0) stars from the literature available at the time. 
Their findings revealed that 11 out of 42 ($\sim$26\%) known UMP stars remain confined within 3 kpc of the Milky Way plane throughout their orbital lifetime. Moreover, 10 out of these 11 UMP stars exhibit prograde orbits (v$_\phi$ > 0). 
Building on this work, \citet{2020sestito} arrived at a similar conclusion using a significantly larger dataset comprising 1027 V/EMP stars. 
This expanded dataset incorporated spectroscopic follow-up data from the Pristine survey \citep{2019aguado} and LAMOST spectroscopy \citep{2018li}. 
Their analysis revealed that approximately 31\% of stars with |z| < 3 kpc observed today never venture beyond 3 kpc of the disc plane. 
Additionally, they examined their sample in the J$_\phi$-J$_z$ projection of the action space. \citet{2020sestito} observed that the number of stars in prograde disc-like orbits (characterised by high J$_\phi$ and low J$_z$) is significantly greater than the number of stars in retrograde disc-like orbits, with a statistical significance of 5$\sigma$. 

\begin{figure}
    \centering
    \includegraphics[width=\columnwidth]{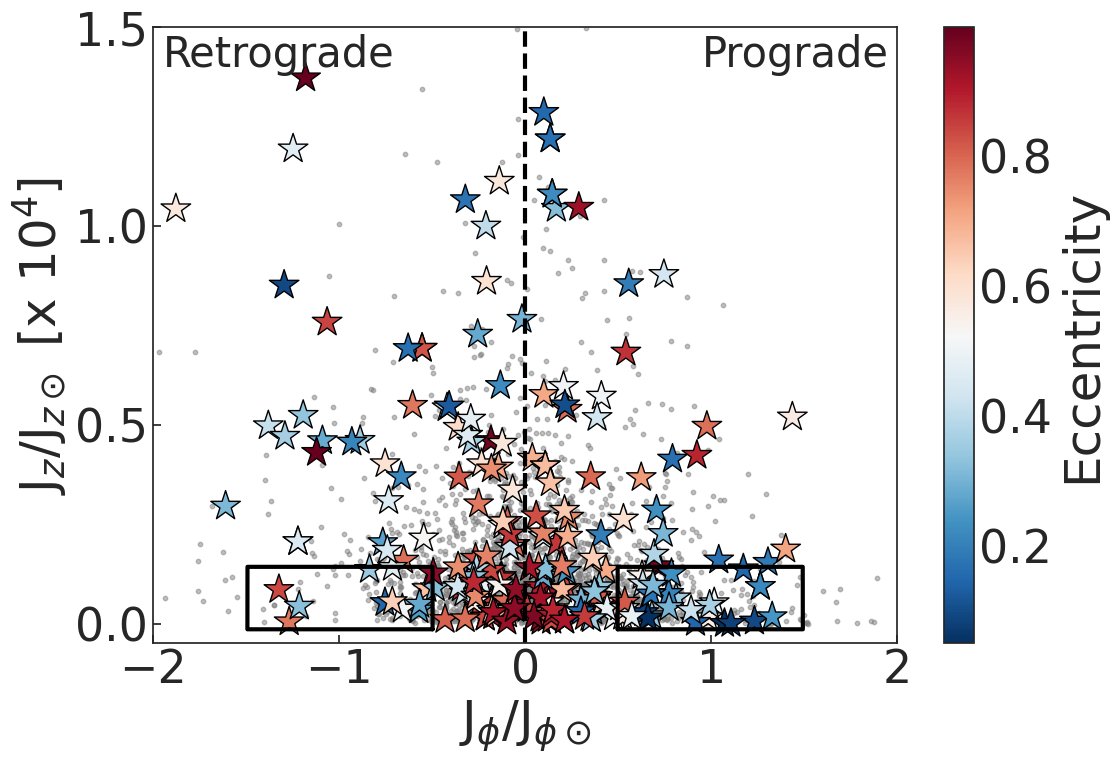}
    \caption{Distribution of low-metallicity stars from the INT sample in the J$_\phi$--J$_z$ action space. The action quantities are scaled by the solar values (i.e., J$_{\phi\odot}$ = 2009.92 kpc km/s, J$_{z\odot}$ = 0.35 kpc km/s). The two black boxes represent the prograde and the retrograde disc-like orbits showing an overdensity in the prograde direction. All EMP stars observable by the spectroscopic follow-up program at INT are shown in grey.}
    \label{planar}%
\end{figure}

We conducted a similar orbital analysis to \citet{2019sestito,2020sestito} and investigated the behavior of our V/EMP stars in the action space, as depicted in Figure \ref{planar}.
This figure can be directly compared to the results from \citet{2020sestito}. We see an overdensity of prograde V/EMP stars confined to the plane up to and even higher than the solar azimuthal action with low to intermediate eccentricities.
Our findings indicate that the fractions of VMP and [Fe/H] < -2.5 (chosen instead of EMP due to low number statistics in EMP) stars with present-day |z| < 3 kpc and Z$_{max}$ < 3 kpc are 31\% and 21\%, respectively, which align very closely with the results of \citet{2020sestito}. 
In the rotation versus vertical action plane, we identified stars within the black dashed boxes representing disc-like prograde and retrograde orbits. 
The region corresponding to low-Z$_{max}$ corotating stars is denser by 4$\sigma$ (entire sample), 3.5$\sigma$ (VMP) and 3.8$\sigma$ ([Fe/H] < -2.5) compared to its retrograde counterpart (using Poisson statistics), slightly lower than the findings reported by \citet{2020sestito}, but still significantly more prograde than retrograde. 
It is essential to note, however, that the methodologies, sample selections, and selection functions employed in these studies vary significantly. The action space distribution of these confined, confined planar (from section \ref{4.2}) and prograde planar (disc-like) stars are shown in Appendix \ref{c}.

\citet{2023zhang} investigated this with XGBOOST derived metallicities from XP spectrophotometry and found a similar (if not higher significance) prograde to retrograde ratio of stars in the metallicity range -1.6 < [M/H] < -3. 
They suggest that this asymmetry can be explained by a combination of a stationary and a prograde halo (proto-MW) component in the VMP end. 
In a recent work by \citet{2024ardernarentsen}, that investigates orbits of VMP stars in the inner Galaxy (R < 3.5 kpc), they report that VMP stars are likely consisting of a mix of a prograde and a stationary halo component. The apocentre distribution shows that VMP stars are very centrally concentrated, with a steep drop towards higher apocentre (and not many with apocentre > 10 kpc). The apocentre and Z$_{max}$ distribution of our planar stars are shown in Appendix \ref{c}.

The chemical properties of these planar stars indicate that a variety of different systems contribute to the formation of this population. \citet{2024dovgal} shows that the large scatter in $\alpha$-elements of the planar star might be indicative that the prograde planar stars is composed of multiple systems. \citet{2024sestito} shows that a planar prograde star with high eccentricity has the same chemical signatures of ultra-faint dwarf stars. Spectroscopic follow-up of some planar stars (retrograde+prograde) show that these observed stars formed in the same formation site, which look like a classical dwarf galaxy \citep{2024sestitoloki}. 
Further investigation, incorporating age distributions and/or more extensive chemical abundance measurements along with extensive orbital analysis in the V/EMP end, is essential to definitively determine the exact metallicity threshold at which the early Milky Way disc originated \citep{2024viswanathanspinup}.
Stars from this spectroscopic follow-up and the ongoing follow-up will add more information in light of this problem and can be potential targets for high-resolution follow-up in the future.

\subsection{C-19 stellar stream new members}

\begin{figure}
    \centering
    \includegraphics[width=\columnwidth]{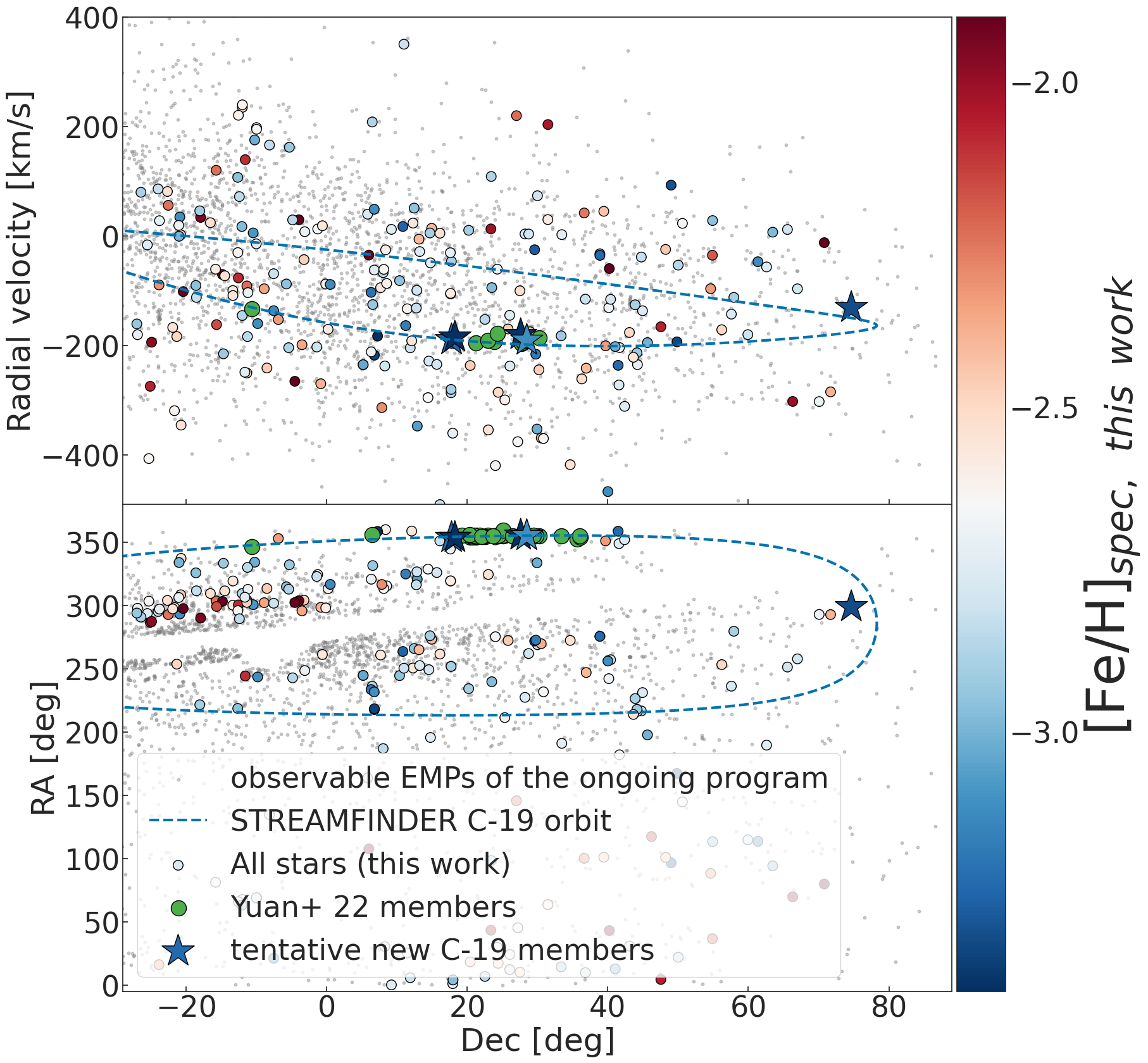}
    \caption{C-19 stellar stream members from this work. The declination vs radial velocity distribution (top) and declination vs right ascension distribution (bottom) of the INT sample of V/EMP stars are  colour-coded by their inferred spectroscopic metallicities with the C-19 orbit from the \texttt{STREAMFINDER} algorithm and the \citet{2022yuan} identified C-19 members overlaid. The tentative C-19 members identified in this work are shown as large star markers. All EMP stars observable by the spectroscopic follow-up program at INT are shown in grey.}
    \label{c19-2}%
\end{figure}

\begin{figure}
    \centering
    \includegraphics[width=\columnwidth]{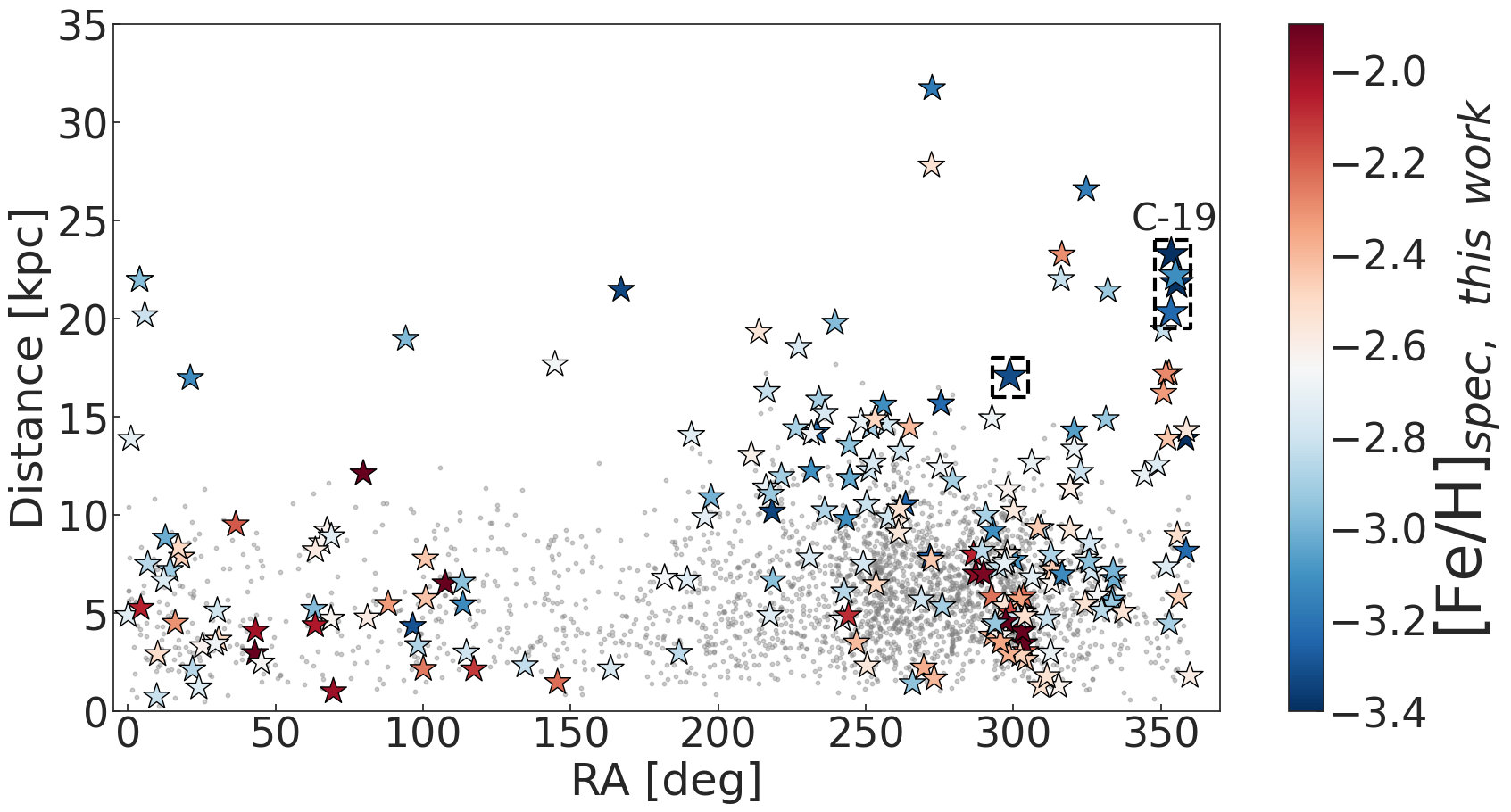}
    \caption{Right ascension vs distance distribution of V/EMP stars from this work colour-coded by their spectroscopic metallicities. The identified C-19 members are shown inside black dashed boxes. All EMP stars observable by the spectroscopic follow-up program at INT are shown in grey.}
    \label{c19-1}%
\end{figure}

The discovery of the most metal-poor stream C-19 by \citet{2021ibata,2022martin,2022martinb} has sparked considerable interest, with its interpretation as a disrupted Globular Cluster (GC) based on several key observations. C-19 stream is the first and foremost exceptionally low-metallicity structure in the universe. Notably, its remarkably narrow metallicity dispersion and a range in Na abundances are consistent with observations of globular clusters \citep{2022martinb,2022yuan}, and contrasting with observations in dwarf galaxies or among the Galactic halo field stars. What sets C-19 apart is its mean metallicity ([Fe/H] = -3.4, EMP), which surpasses that of even the most metal-poor GC \citep{2021vasiliev} and represents the lowest metallicity structure ever found. The absence of Galactic Globular Clusters observed below a metallicity of -2.5 has led to interpretations suggesting a metallicity floor beyond which GCs cannot form. 
The Phoenix stream \citep[mean $\text{[Fe/H]}$ = -2.7,][]{2016balbinot,2020wan} and ED-2 stream \citep[mean $\text{[Fe/H]}$ = -2.6,][]{2023balbinot,2024balbinot} are other streams that challenge the GC metallicity floor. 
However, despite chemical evidence suggesting a Globular Cluster origin, the dynamical width and velocity dispersion of C-19 resemble those of a dwarf galaxy. One hypothesis suggests that the stream may have originated from a dark matter-dominated dwarf galaxy, with the star cluster comprising most of the mass in the system, although the remnants of such a dwarf galaxy have yet to be discovered \citep{2022errani}. Alternatively, \citet{1999helmi} demonstrated through numerical simulations that the velocity dispersion of streams can increase significantly at the turning point of their orbit, potentially explaining C-19's large velocity dispersion. However, determining whether the stream is indeed at its turning point poses a challenge due to its sensitivity to the assumed potential.
Thereby, finding more members on C-19's orbit and brightest members (better suitable for high-resolution follow-up) that belong to the streams will help us constrain the origin of this ancient structure. 

Figure \ref{c19-2} shows the right ascension versus declination in the bottom panel and radial velocity versus declination in the top panel for all the stars analysed in this work, colour-coded by the inferred spectroscopic metallicity. 
Overlaid are the C-19 members from \citet{2022yuan} in green, with one radial velocity member discovered 30$^{\circ}$ away from the main body of the stream. 
From the metallicities and the STREAMFINDER inferred C-19 orbit, we find five (three new, two rediscovered) C-19 members (inferred metallicities between -3.24 and -3.6, all consistent with C-19 mean metallicity within 2$\sigma$ considering the measurement and systematic errors), two of which are the brightest members of the stream discovered yet at the tip of the red giant branch (RGB).
One of the members discovered is 50$^\circ$ away from the main stream counterpart, above the disc plane at positive height above the plane. This is the first C-19 member discovered above the disc plane. 
This member increases the span of the stream on the sky. 
The overdensity of stars in one part of the stream could be due to the coverage in the region of the Pristine survey as it probes deeper to find C-19 members with photometric metallicities. 
The mean metallicity of the stream with the new members is -3.33$\pm$0.07, the stream width is 0.39$^\circ\pm$0.07$^\circ$ and the velocity dispersion (excluding the member above the disc plane) is 4.11$\pm$2.05 km/s consistent with the literature values. 
We note that C-19 member stars are not associated with any of the accretion events in Figure \ref{kinematics}, according to the literature selections. Although, \citet{2022malhan} suggests that C-19 is mostly likely a stream associated with the LMS-1/Wukong structure.
C-19 stream and LMS-1/Wukong have different orbits. Even though they are both polar, the orbital planes are separated by a large angle.
A more comprehensive look at the members of C-19 all over the sky with chemistry and dynamics will be presented in the upcoming work (Yuan et al., in prep.) which will provide more insight into the progenitor and formation of the C-19 stream. 

Figure \ref{c19-1} shows the distance versus right ascension plane of all the stars observed and analysed in this work, colour-coded by metallicity. 
Tentative C-19 members from this work are highlighted with a dashed black box and the members from the main body of the stream are clustered at around 21.5 kpc, while the member above the disc plane lies at 17 kpc which indicates a distance gradient across the stream.
The mean distance of the stream is slightly higher than what has been reported in the literature. 
\citet{2021ibata} estimated the distance to C-19 using the \texttt{STREAMFINDER} algorithm, which determines a most likely heliocentric distance, based on the local distribution of stars on the
sky, in proper motion space, and in CMD space which is not individually precise, but for a stream as a whole, it is a good first guess. They found a distance between 16 and 22 kpc with no distance gradient along the stream. \citet{2022martinb} estimates the distance to C-19 using 7 blue horizontal branch stars (17.5 kpc) at a favoured distance
of 18 kpc to minimise the difference between the FeI and FeII iron abundances, and a distance estimate based on \textit{Gaia} parallaxes and the best fit C-19 orbit (20.9 kpc).
All these stars have bad parallax, which means that the PDF of the distance inference depends highly on the photometry, extinction, systematic offset in the method and the choice of isochrone which is most likely a bad match for [Fe/H]<-3 stars. Therefore, we do not expect these stars to have distances as good as standard candles.
However, \citet{2024bonifacio}. followed-up and analysed fainter sub-giant branch stars from the C-19 stream and found that an EMP isochrone fits better with 20.9 kpc distance in the subgiant branch that would match well with our estimate.  
Based on our analysis, we find a mean distance of 21.5 kpc, which is a slightly larger distance than the literature-reported distances. 

In Figure \ref{c19-1}, the INT observable stars are in grey. Because these are good parallax stars, we can see that the grey points end at about 10 kpc while due to the bad parallax selected giants, our sample goes up to 35 kpc in the EMP bright end, much farther than what is possible using only good parallax selections. 
Given the success demonstrated in this follow-up paper, we pave the way for the use of bright and
distant metal-poor stars and streams from these photometric studies as invaluable guides to the early universe with a more homogeneous selection of giant stars \citep{2024viswanathangiants}.

\section{Conclusions and outlook}\label{6}

Thanks to the recent \textit{Gaia} DR3, we have spectro-photometric information from the XP spectra for $\sim$219 million stars. Using these data, \citetalias{2023martin} calculated synthetic, narrowband, metallicity-sensitive \textit{CaHK} magnitudes that mimic the observations of the Pristine survey, and infer photometric metallicities using the narrowband \textit{CaHK} and broadband \textit{Gaia} magnitudes trained on the survey's training sample. In this work, we presented the first dedicated spectroscopic follow-up of V/EMP stars from this all-sky catalogue of photometric metallicities. 
The target selection was based on the quality cuts recommended by the catalogue paper, \citetalias{2023martin}, at the bright end (Figure \ref{target-selection} and \ref{allsky-targets}).
We used the INT/IDS instrument in La Palma for the low- to medium-resolution spectroscopic follow-up in the calcium triplet region, which is widely used to study VMP stars (Table \ref{telescope}).
We inferred radial velocities and spectroscopic metallicities using the pipeline based on \citet[][Figure \ref{spectra}]{2024viswanathan}. 
Due to the available 6D phase space information, we also took a deeper look into the dynamics of these stars in the context of Galactic archaeology.
Our main results are presented as follows:

\begin{table*}
\caption{Description of the columns of the INT follow-up spectra of 215 V/EMP stars as a catalogue made available publicly in this work.} 
\label{catalogue}
\centering
\begin{tabular}{llll}
\hline \hline
 Column & Description & Unit & Type \\
 \hline
source\_id & \textit{Gaia} DR3 Source ID & unitless & longint \\
ra & \textit{Gaia} DR3 right ascension in ICRS (J2016) format & degrees & float \\
dec & \textit{Gaia} DR3 declination in ICRS (J2016) format & degrees & float \\
G\_0 & de-reddened \textit{Gaia} G magnitude & unitless & float\\
BP\_0 & de-reddened \textit{Gaia} G$_{BP}$ magnitude & unitless & float\\
RP\_0 & de-reddened \textit{Gaia} G$_{RP}$ magnitude & unitless & float\\
distance & Photometric distance or inverted parallax based distance derived in this work & mas & float\\
distance\_error & Uncertainty on the photometric distance & mas & float\\
snr & Signal-to-noise ratio of the normalised INT CaT spectra & unitless & float \\
v\_r & Radial velocity measured from the INT CaT spectra & km/s & float \\
dv\_r & Uncertainty in the measured radial velocity & km/s & float \\
ew1 & Equivalent width of the first calcium triplet line around 8498 \AA & \AA & float \\
dew1 & Uncertainty on the equivalent width of the first calcium triplet line around 8498 \AA & \AA & float \\
ew2 & Equivalent width of the second calcium triplet line around 8542 \AA & \AA & float \\
dew2 & Uncertainty on the equivalent width of the second calcium triplet line around 8542 \AA & \AA & float \\
ew3 & Equivalent width of the third calcium triplet line around 8542 \AA & \AA & float \\
dew3 & Uncertainty on the equivalent width of the third calcium triplet line around 8662 \AA & \AA & float \\
feh & Spectroscopic metallicity derived in this work using INT CaT spectra & unitless & float \\
dfeh & Measurement uncertainty associated with the spectroscopic metallicity derived & unitless & float \\
E & Total energy & kpc km$^2$/s$^2$ & float \\
Jphi & rotational action vector & kpc km/s & float \\
Jz & vertical action vector & kpc km/s & float \\
Jr & radial action vector & kpc km/s & float \\
zmax & maximum height above the plane of the orbit & kpc & float \\
rapo & apocentre of the orbit & kpc & float \\
rperi & pericentre of the orbit & kpc & float \\
ecc & eccentricity of the orbit & unitless & float \\
group\_number & Group number for stars that are clustered in IOM space & unitless & int \\
\hline
\end{tabular}
\tablefoot{The reduced 1D spectra from this work is also made available publicly. All these data are available at the CDS}
\end{table*}

\begin{itemize}
    \item The spectroscopic metallicities are inferred for 215 stars at a precision of 0.08 in measurement added in quadrature with 0.15 dex in systematic uncertainty down to $\sim$-3.6 in metallicity (Figure \ref{gmag-feh-distribution}).
    \item The photometric metallicities from the Pristine-\textit{Gaia} synthetic catalogues agree well with the inferred spectroscopic metallicities from our follow-up with $\sim$77\% and $\sim$38\% success rates of having [Fe/H] < -2.5 and -3.0 stars, respectively. The outliers ([Fe/H > -2.0) are reduced to 3\% with no catastrophic outliers ([Fe/H > -1.0). This is a huge improvement over the existing methods that search for EMP stars (Figure \ref{feh-comparison} and Table \ref{comp}). With the enhanced success rates observed, we anticipate that the photometric metallicities obtained from the Pristine survey and the Pristine-\textit{Gaia} synthetic catalogues will facilitate the identification of more than 10,000-20,000 uniformly analsed EMP stars through the WEAVE low-resolution (R$\sim$5000) follow-up of Pristine EMP stars \citep{2024weave}. This will enable unprecedented statistical analyses of the metal-poor Galaxy.
    \item From the kinematics of the V/EMP stars from this work, we find 20\% are in confined orbits, 46\% in the inner halo, and 34\% in the outer halo orbits (see panels c, e, and g of Figure \ref{kinematics}).
    \item We associate V/EMP stars with several known accretion events such as GES, LMS-1/Wukong, Thamnos, Sequoia, Sagittarius, and Helmi streams, which would define the EMP end of their metallicity distribution. Some of the small and tight clumps in the E-L$_z$ space could belong to low-mass mergers in the distant past (see panel a, and panels b, d, and f of Figure \ref{kinematics}).
    \item We find that 31\% of our VMP stars with |z| < 3 kpc do not go beyond 3 kpc in their orbital history. Along this line, the prograde region with low vertical action (similar to disc-like orbits) is 4$\sigma$ overdense compared to the retrograde counterpart, suggesting that an important fraction of V/EMP stars reside in the disc plane (Figure \ref{planar}).
    \item We associate five stars from our follow-up with the most metal-poor stellar stream C-19, three of which are new. One of these stars is 50$^\circ$ from the main body of the stream, and is the first star belonging to C-19 with positive height above the disc plane. Adding these bright members of C-19 helps to improve the constraints on its progenitor and orbital history. We suggest that the stream might lie farther out than has been reported in the literature (mean distance of about 21.5 kpc, compared to $\sim$18 kpc reported in \citealt{2022martinb}), while the mean metallicity, velocity dispersion and dynamical width are consistent with the literature values (Figure \ref{c19-2} and \ref{c19-1}).
\end{itemize}

With this follow-up, we characterise the success rates of the Pristine-\textit{Gaia} synthetic catalogue of photometric metallicities and the implications of V/EMP stars from a chemokinematics point of view. Future medium- to high-resolution follow-up with chemical abundances and large multi-object spectrographs such as WEAVE \citep{2024weave} and 4MOST \citep{20194most} will take the full chemodynamical analysis of V/EMP stars further, allowing us to study some of the smallest and earliest galaxies that merged into the Milky Way. 

\begin{acknowledgements}
      We express our gratitude to the reviewer for dedicating their valuable time and providing insightful contributions that significantly enhanced the quality of our manuscript along with their kind words about this work. AV thanks Ewoud Wempe and Emma Dodd for helpful discussions. ZY, AAA, NFM, RAI, and SR gratefully acknowledge support from the French National Research Agency (ANR) funded project "Pristine" (ANR-18-CE31-0017) along with funding from the European Research Council (ERC) under the European Unions Horizon 2020 research and innovation programme (grant agreement No. 834148). AAA acknowledges support from the Herchel Smith Fellowship at the University of Cambridge and a Fitzwilliam College research fellowship supported by the Isaac Newton Trust. ES, MB, and MM acknowledge funding through VIDI grant "Pushing Galactic Archaeology to its limits" (with project number VI.Vidi.193.093) which is funded by the Dutch Research Council (NWO). TM was supported by a Gliese Fellowship at the Zentrum f\"{u}r Astronomie, University of Heidelberg, Germany. Research by ADD is supported by NSF grant AST-1814208. JMA acknowledges support from the Agencia Estatal de Investigaci\'on del Ministerio de Ciencia en Innovaci\'on (AEIMICIN) and the European Social Fund (ESF+) under grant PRE2021-100638. CAP acknowledge financial support from the Spanish Ministry of Science and Innovation (MICINN) project PID2020-117493GB-I00. JMA acknowledges support from the Agencia Estatal de Investigaci\'on del Ministerio de Ciencia en Innovaci\'on (AEI-MICIN) and the European Regional Development Fund (ERDF) under grant number AYA2017-89076-P, the AEI under grant number CEX2019-000920-S and the AEI-MICIN under grant number PID2020-118778GBI00/10.13039/501100011033. DA also acknowledges financial support from the Spanish Ministry of Science and Innovation (MICINN) under the 2021 Ram\'on y Cajal program MICINN RYC2021-032609.  Based on observations obtained with MegaPrime/MegaCam, a joint project of CFHT and CEA/DAPNIA, at the Canada-France-Hawaii Telescope (CFHT) which is operated by the National Research Council (NRC) of Canada, the Institut National des Sciences de l'Univers of the Centre National de la Recherche Scientifique of France, and the University of Hawaii. This work has made use of data from the European Space Agency (ESA) mission Gaia (https://www.cosmos.esa.int/gaia), processed by the Gaia Data Processing and Analysis Consortium (DPAC, https://www.cosmos.esa.int/web/gaia/dpac/consortium). Funding for the DPAC has been provided by national institutions, in particular the institutions participating in the Gaia Multilateral Agreement. We are deeply grateful to the INT support astronomers, Rosa Hoogenboom, and Cl\'ar-Br\'id Tohill, for carrying out observation in service mode during Covid-19 times. We also thank the INT staff for answering all our questions both administrative and scientific during the observing runs. The INT/IDS is operated on the island of La Palma by the Isaac Newton Group of Telescopes in the Spanish Observatorio del Roque de los Muchachos of the Instituto de Astrof\'isica de Canarias. This research was supported by the International Space Science Institute (ISSI) in Bern, through ISSI International Team project 540 (The Early Milky Way). This research has been partially funded from a Spinoza award by NWO (SPI 78-411). AV also thanks the availability of the following packages and tools that made this work possible: \texttt{vaex} \citep{2018vaex}, \texttt{pandas} \citep{2022pandas}, \texttt{astropy} \citep{2022astropy}, \texttt{NumPy} \citep{2006numpy,2011numpy}, \texttt{SciPy} \citep{2001scipy}, \texttt{matplotlib} \citep{2007matplotlib}, \texttt{seaborn} \citep{2016seaborn}, \texttt{agama} \citep{2019vasiliev}, \texttt{gala} \citep{2017pricewhelan}, \texttt{galpy} \citep{2016jupyter}, \texttt{pyraf} \citep{2001pyraf}, \texttt{iraf} \citep{1999iraf}, \texttt{gaiadr3-zeropoint} \citep{2021lindegreen}, \texttt{JupyterLab} \citep{2016jupyter}, and \texttt{topcat} \citep{2018topcat}.

\end{acknowledgements}

\bibliographystyle{mnras}
\bibliography{aanda}

\begin{appendix}

\section{Galactic distribution of V/EMP stars}\label{a}

\begin{figure*}
    \centering
    \includegraphics[width=\textwidth]{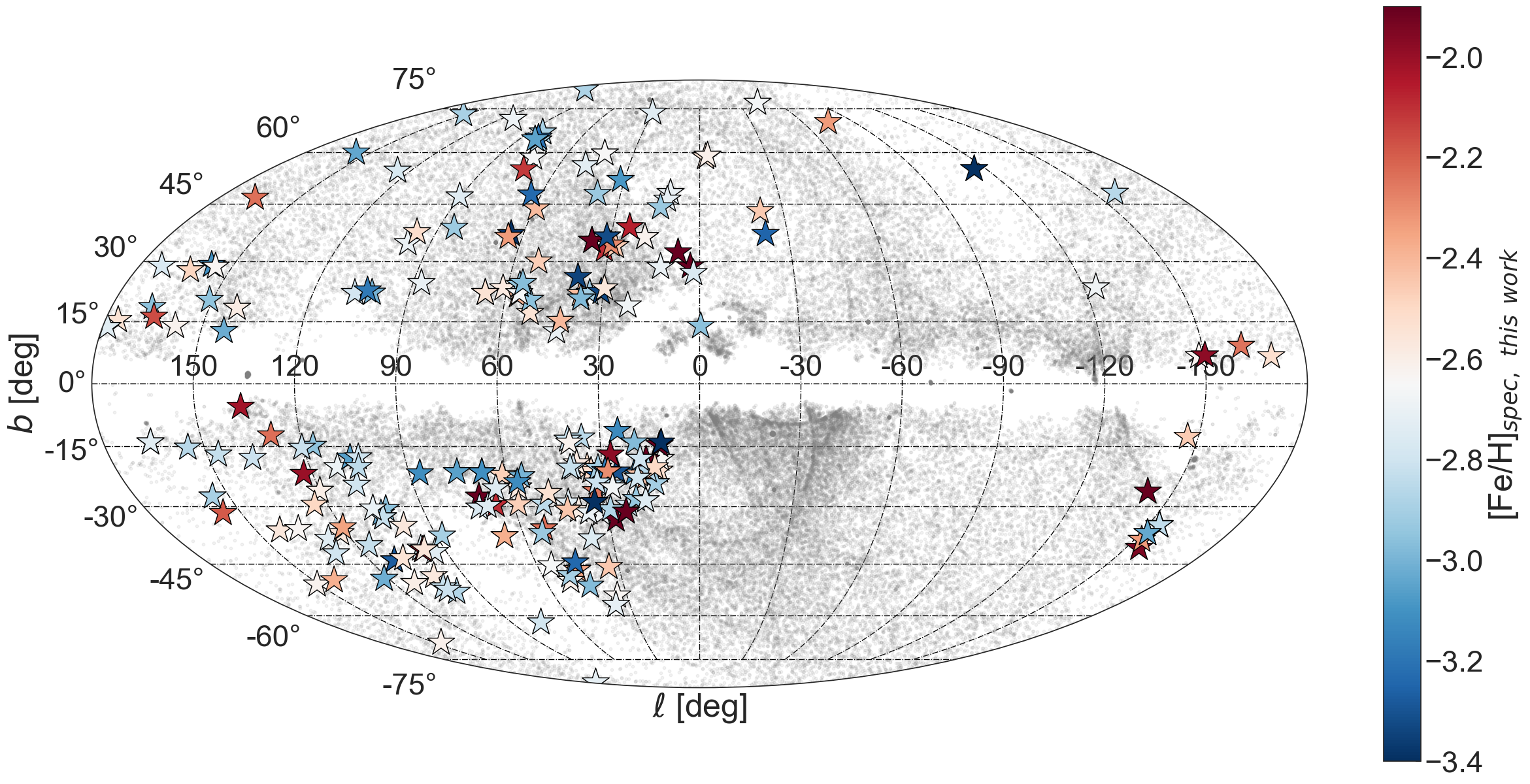}
    \begin{flushleft}
        \includegraphics[width=0.86\textwidth]{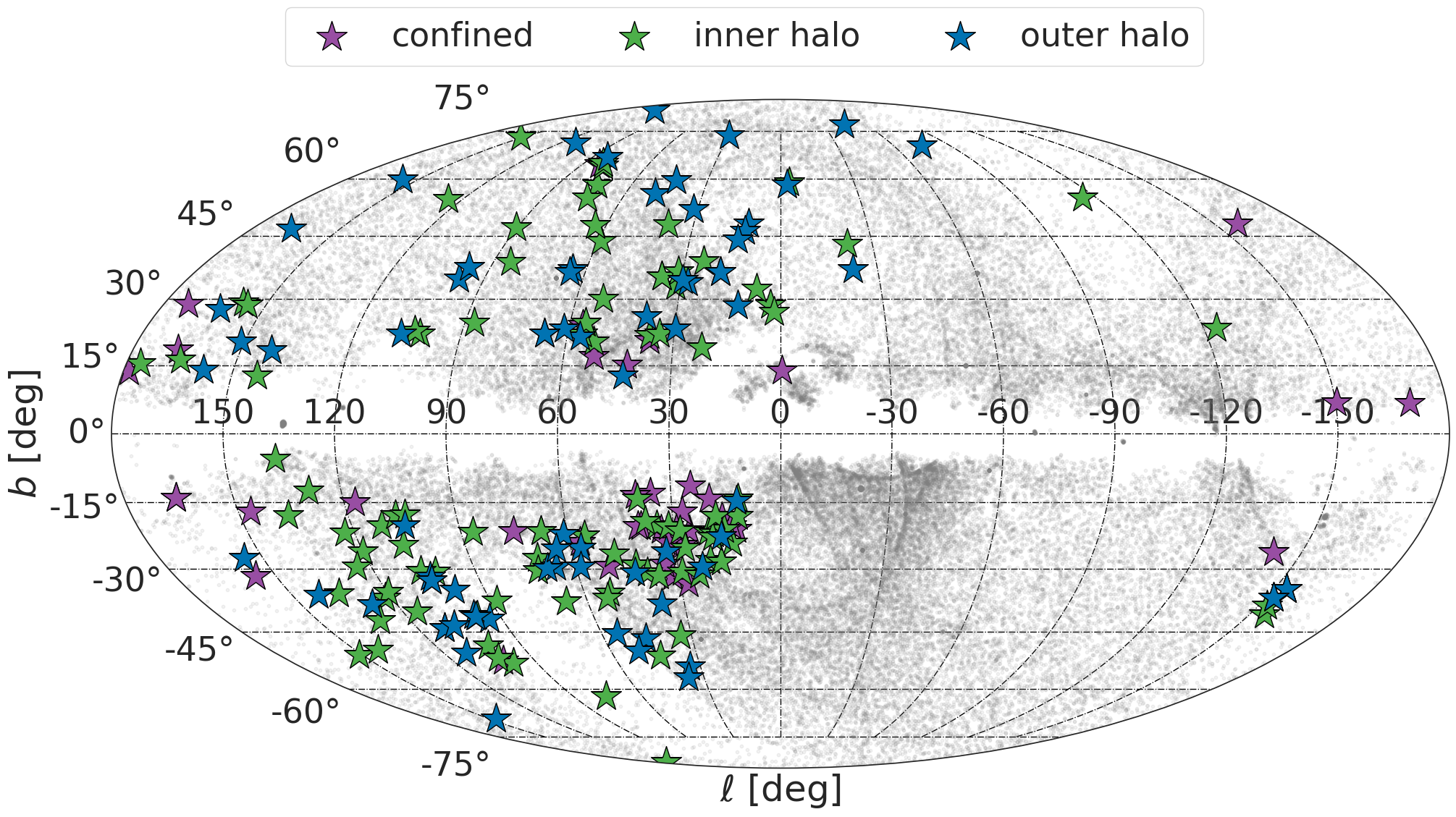}
    \end{flushleft}
    \caption{Mollweide projection of the Galactic coordinates for the V/EMP stars with [Fe/H]<-2.5 in the Pristine-\textit{Gaia} synthetic catalogue (in grey). The large-scale patterns visible in the map are mainly attributed to the scanning law of the \textit{Gaia} satellite. The coloured stars represent the subset of stars observed and analysed in this spectroscopic follow-up, colour-coded by their inferred spectroscopic metallicities (top). The different coloured stars belong to confined, inner, and outer halo stars, as defined in section \ref{4.2} (bottom).}
    \label{app-sky}%
\end{figure*}

Figure \ref{app-sky} top panel shows the sky distribution of stars spectroscopically followed-up in this work colour-coded by their inferred spectroscopic metallicities from this work. We do not see any striking feature or substructure(s) in this view. The distribution of both VMP and EMP stars are almost random on the sky showing that the Galactic position does not affect our inferred metallicities. The bottom panel of Figure \ref{app-sky} shows the sky distribution of confined to the disc plane, inner halo and outer halo stars as defined in section \ref{4.2} in the same colour scheme. We can see that the confined stars are almost within 30$^\circ$ of the Galactic latitude, while the inner and outer halo stars are distributed evenly over the sky.

\FloatBarrier

\section{Additional chemokinematics of the V/EMP stars}\label{b}

\begin{figure}[htbp!]
    \centering
    \includegraphics[width=0.995\columnwidth]{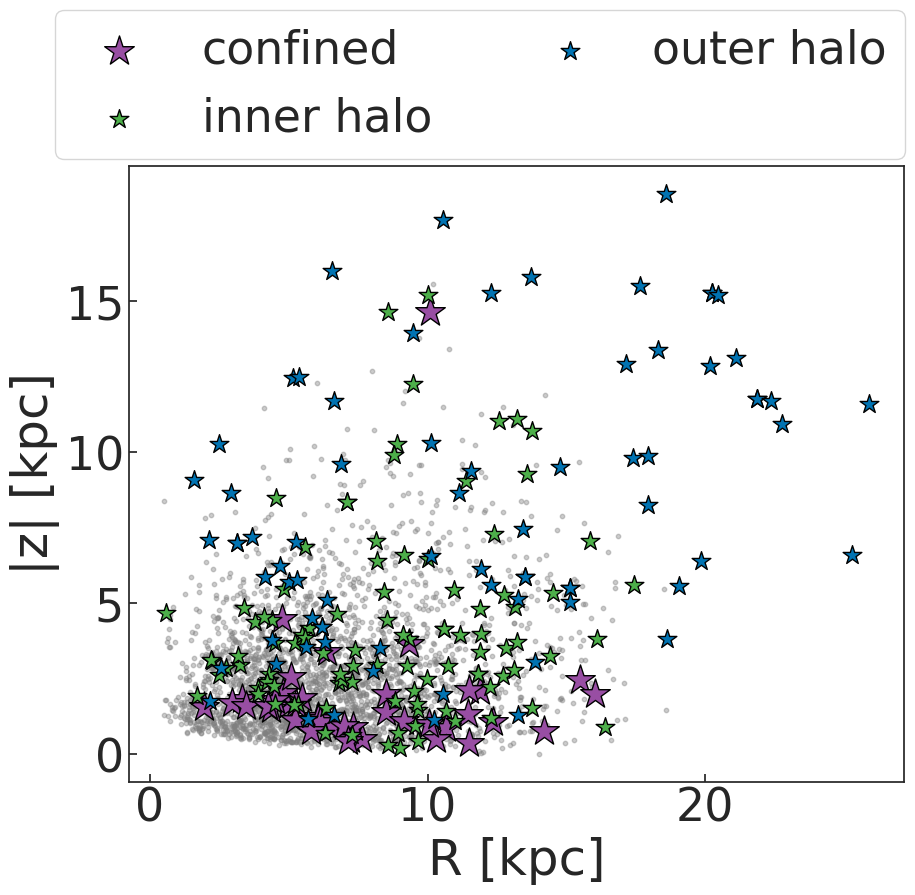}
    \includegraphics[width=\columnwidth]{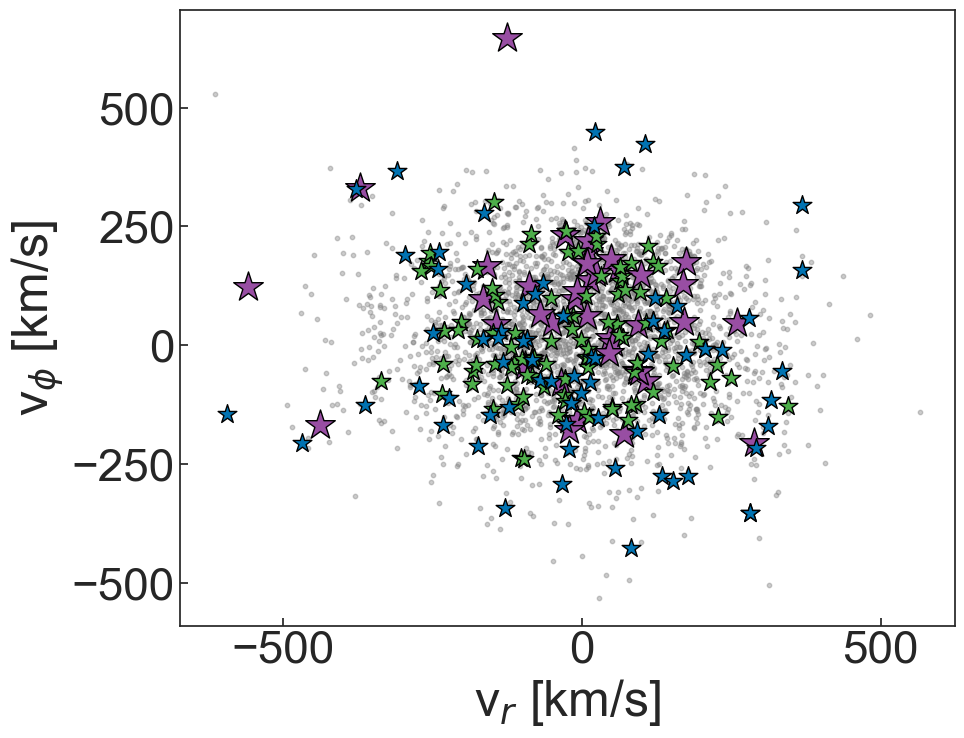}
    \includegraphics[width=0.995\columnwidth]{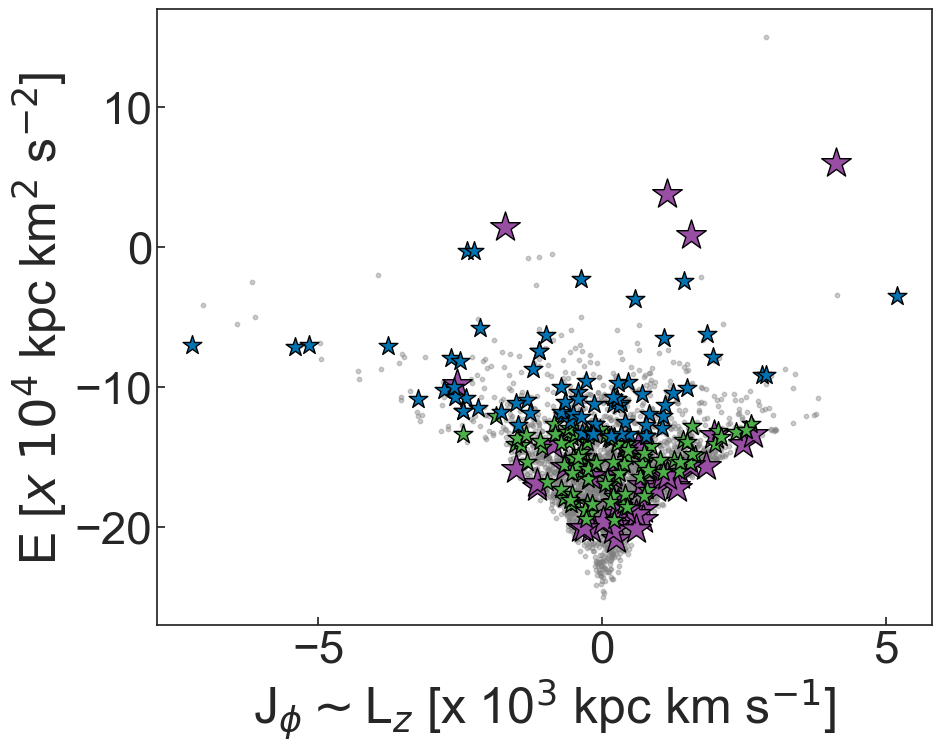}
    \caption{(Top) Galactocentric radius vs absolute scale height distribution of confined, inner and outer halo stars as defined in section \ref{4.2}. (Middle) Velocity distribution of these stars in v$_\phi$ vs v$_r$ space. (Bottom) Energy vs angular momentum in z-direction for the classified stars. All EMP stars observable by the spectroscopic follow-up program are shown in grey.}
    \label{app-kine}
\end{figure}

In Figure \ref{app-kine}, we additionally show Galactic distribution and kinematics view of the confined to the plane, inner halo and outer halo group defined in section \ref{4.2}. The top panel of Figure \ref{app-kine} shows the radius ($\sqrt{x^2+y^2}$) versus absolute scale height distribution of our observed V/EMP stars. We see that most of the confined stars have a flattened distribution over the z-plane and are solar-neighbourhood stars with more stars towards the inner Galaxy, while the inner and outer halo stars go out to 15-20 kpc in radius and scale height as expected. The middle panel of Figure \ref{app-kine} shows the velocity distribution in azimuth versus radial direction. The overall distribution looks quite isotropic with a small prograde rotation (about 20 km/s). The confined stars peak at around 180 km/s with the high-energy stars going out to about 600 km/s. Inner halo stars are have a smaller spread in their velocities compared to outer halo, as expected. The bottom panel of Figure \ref{app-kine} shows energy versus angular momentum in the z-direction (IOM space) for the confined, inner and outer halo stars. We see a clear difference in the energy distribution between inner and outer halo stars with inner halo stars sinked deeper into the potential than the outer halo stars. Confined stars have a mix of prograde and retrograde orbits. One other interesting note is that all our high-energy (unbound) stars belong to the confined plane suggesting a very high apocentre and small Z$_{max}$, which favours the argument that they are less likely to be hypervelocity stars but their extreme properties are more likely due to uncertainties in positions and velocities. 

\FloatBarrier

\section{Planar metal-poor stars in other kinematic spaces}\label{c}

\begin{figure}[h!!!!!]
    \centering
    \includegraphics[width=\columnwidth]{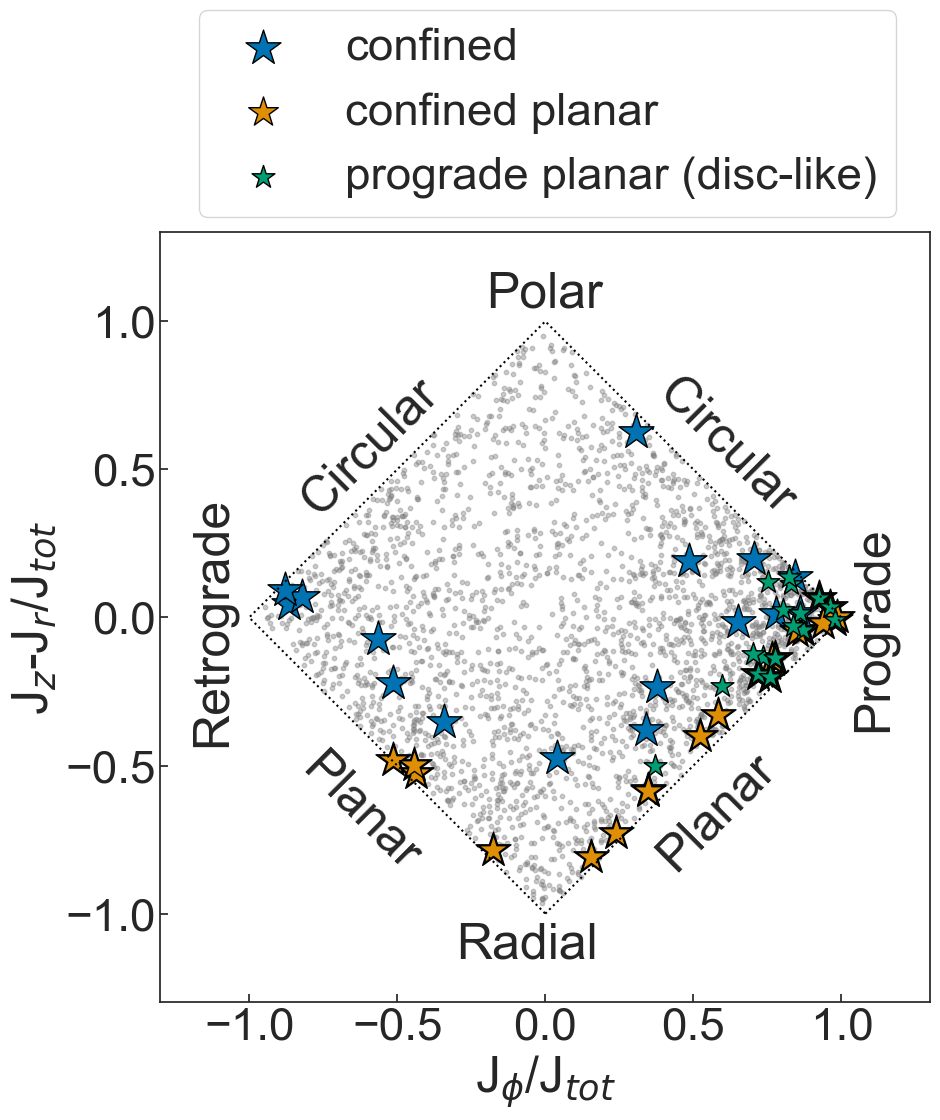}
    \includegraphics[width=\columnwidth]{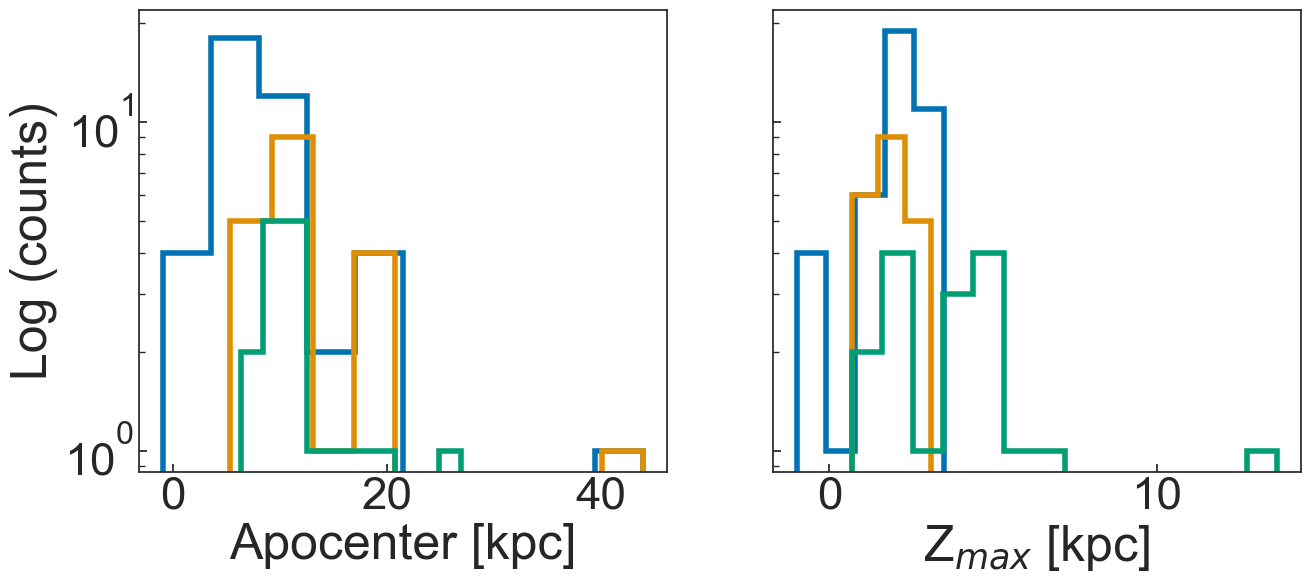}
    \caption{(Top) Action space diamond, the difference between the vertical and radial component of the action, vs the rotational component of the action with the axes normalised by J$_{tot}$ = |J$_\phi$ | + J$_r$ + J$_z$ for confined and confined planar stars as defined in section \ref{4.2} and prograde planar (disc-like) stars, as defined in section \ref{5.1}. All EMP stars observable by the spectroscopic follow-up program are shown in grey (Bottom) Apocentre and Z$_{max}$ distribution of these stars are shown in the same colour.}
    \label{app-planar}
\end{figure}
\FloatBarrier

In Figure \ref{app-planar}, we show action diamond space (normalised to the sum of the three actions of
motion) in the top panel and apocentre and Z$_{max}$ distribution of confined (Z$_{max}$ < 3.5 kpc and apocentre > 3.5 kpc) and confined planar (confined + Z$_{max}$/apocentre < 0.2) stars from section \ref{4.2} and prograde planar (disc-like, prograde box from Figure \ref{planar}) stars from section \ref{5.1} in the bottom panel. We can see that the confined stars are close to the plane, but are not always in the plane (few circular and radial stars), while confined planar stars are in the plane (prograde or retrograde). We see that the prograde planar (disc-like) stars is approximately around the same region of overdensity seen with metal-poor stars moving close to the plane in the action diamond (prograde and planar) from various spectroscopic surveys as shown in \citet{2024alvar}. The apocentre distribution is close to the solar neighbourhood and falls off immediately after, favouring the scenario that these stars could originate from a centrally concentrated, slightly prograde proto-Galaxy as shown by \citet{2023zhang,2024ardernarentsen}. However, it is important to note that our sample size for this population is too small to say anything conclusive.    

\end{appendix}

\end{document}